\documentclass[fleqn,usenatbib,useAMS]{mnras}
\pdfoutput=1 
\usepackage{graphicx}	
\usepackage{amsmath}	
\usepackage{amssymb}	
\usepackage{multicol}        
\usepackage{bm}		
\usepackage{microtype}
\usepackage{lineno}

\newcommand{\bs}[1]{\boldsymbol{#1}}
\newcommand{\given}{\,|\,}
\newcommand{\norm}{\mathcal{N}}
\newcommand{\dd}{\mathrm{d}}

\newcommand{\mean}[1]{\left< #1 \right>}

\newcommand{\truepos}{\ensuremath{\tilde{\bs{x}}}}
\newcommand{\gaia}{\emph{Gaia}}

\title{Three-dimensional structure of the Sagittarius dSph core from RR Lyrae}

\author[P. Ferguson \& L. Strigari]{Peter S. Ferguson$^{1,2}$\thanks{Contact e-mail: \href{mailto:petersferguson@tamu.edu}{petersferguson@tamu.edu}}\href{https://orcid.org/0000-0001-6957-1627}{\includegraphics[scale=0.1]{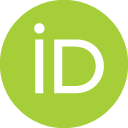}} and Louis E. Strigari\href{https://orcid.org/0000-0001-5672-6079}{\includegraphics[scale=0.1]{orcid.png}}$^{1,2}$
\\
$^{1}$George P. and Cynthia Woods Mitchell Institute for Fundamental Physics and Astronomy, \\Texas A\&M University, College Station, TX 77843, USA\\
$^{2}$Department of Physics and Astronomy, Texas A\&M University, College Station, TX 77843, USA}

\date{}

\pubyear{2020}

\begin{document}
\label{firstpage}
\pagerange{\pageref{firstpage}--\pageref{lastpage}}
\maketitle

\begin{abstract}
    We obtain distances to a sample of RR Lyrae in the central core of the Sagittarius dwarf spheroidal galaxy from OGLE data. We use these distances, along with RR Lyrae from \emph{Gaia} DR2, to measure the shape of the stellar distribution within the central $\sim$ 2 kpc. The best-fit stellar distribution is triaxial, with axis ratios 1 : 0.76 : 0.43. A prolate spheroid model is ruled out at high statistical significance relative to the triaxial model. The major axis is aligned nearly parallel to the sky plane as seen by an Earth-based observer and is nearly perpendicular to the direction of the Galactic center. This result may be compared to cosmological simulations which generally predict that the major axis of the dark matter distribution of subhalos is aligned with the Galactic center. The triaxial structure that we obtain can provide important constraints on the Sagittarius progenitor, as well as the central dark matter distribution under the assumption of dynamical equilibrium.  
\end{abstract}

\section{Introduction}
\par Precise measurements of distances to member stars have provided important information on the three-dimensional structure of the Sagittarius dwarf galaxy and its associated stellar stream. The stream has now been mapped out over the full sky by determining the distances to M-giants~\citep{2003ApJ...599.1082M}, main-sequence, horizontal branch, red giants~\citep{2010ApJ...712..516N,2012ApJ...750...80K,2013ApJ...762....6S,2014MNRAS.437..116B}, and RR Lyrae~\citep{2017ApJ...844L...4S,2017ApJ...850...96H}. These measurements now show that the leading and the trailing arms of the stream extend $\sim 20 -120$ kpc from the main body. The three-dimensional structure of the stream is a necessary input to simulations which attempt to understand its origin~\citep{2010ApJ...714..229L,2011ApJ...727L...2P,2017ApJ...836...92D}. The phase-space structure of the stream may also provide new probes of exotic physics in the dark matter sector~\citep{2006PhRvD..74h3007K,Xu:2019arXiv190408949X}. 

\par In addition to the stream, the structure of the core provides important information on the nature and evolution of Sagittarius.  

The first in-depth photometric and kinematic study by~\cite{1997AJ....113..634I} found a half-light radius for the core of $\sim 1$ kpc, and an average line-of-sight velocity dispersion of $\sim 11$ km/s. Distance estimates to red clump stars indicate that the ratios of the major, intermediate, and minor axes are 1:0.33:0.33. 
The line-of-sight velocity dispersion is now measured out to several half-light radii, and is $\sim 10-15$ km/s, with a cold spot in the central region~\citep{Frinchaboy2012ApJ...756...74F,Majewski2013ApJ...777L..13M}. 

\par Understanding the dynamical structure of both the core and the stream has important implications for constraining the progenitor of Sagittarius and its dark matter properties. It has long been known that the observed geometry of the leading and trailing arms of the streams imply that the progenitor resides in a more massive and extended dark matter halo~\citep{1995ApJ...451..598J,1998ApJ...500..575I}.  Matching the recent kinematic data in the streams implies that the progenitor mass is $\gtrsim 6 \times 10^{10}$ M$_\odot$~\citep{2017MNRAS.464..794G}. Similarly in order to match the kinematics in the core, simulations suggest that the total stellar plus dark matter mass of the progenitor was $\gtrsim 10^{10}$ M$_\odot$~\citep{2010ApJ...725.1516L}. The nature of the progenitor may also be constrained from the lack of rotation signal in the central core~\citep{2011ApJ...727L...2P}. 

\par Measuring the shape and the orientation of dwarf satellite galaxies like Sagittarius is important from the perspective of the $\Lambda$CDM theory of structure formation. Dark matter only simulations of tidally-disrupting satellites find that heavily-stripped subhalos tend to be rounder than those that are less tidally disturbed~\citep{2007ApJ...671.1135K,2015MNRAS.447.1112B}. These simulations also find that the major axes of the subhalos tend to align towards the center of the host dark matter halo. This effect is most pronounced in the outer regions of the subhalo; it is not yet clear how baryons alter both the shapes and the orientations of the subhalos.  

\par The kinematics of the Sagittarius core may be used to determine the dark matter mass distribution in this region. Assuming that the system is in dynamical equilibrium, the mass distribution may be extracted using methods that are typically used on dwarf spheroidal galaxies~\citep{2013NewAR..57...52B}. However, all of these methods are limited because the inputs to them are derived from projected quantities such as the surface density or velocity dispersion.  Since it is the three-dimensional stellar density profile that must be used in the dynamical models, an incorrect input for it may bias the reconstructed luminous and dark mass distributions. For example, in the simplest case of spherically-symmetric stellar distributions, there is a non-unique mapping from projected two-dimensional stellar distribution onto a three-dimensional stellar density distribution. In particular, an observed flat two-dimensional stellar profile may project onto either a central core or a central cusp in three dimensions, and this has important implications for extracting the dark matter distribution~\citep{2010MNRAS.408.2364S}. Distance information on individual stars would provide an important new input to constrain dynamical models~\citep{Richardson:2013hga}.

\par For axisymmetric models, the extraction of the three-dimensional stellar profile from the two-dimensional data is even more difficult. This is because of the projection issue that plagues spherical models, and even more importantly because there is an inclination angle of the major (or minor) axis with respect to the sky plane that must be determined. Only with a measurement of this inclination is it possible to obtain the three-dimensional velocity dispersion from the measured two-dimensional dispersions, and estimate the dark matter distribution (e.g.~\citet{2015ApJ...810...22H}) 

\par Obtaining an empirical three-dimensional stellar distribution requires a precise measurement of the distance to individual member stars. However because of the relatively large distances to a typical dwarf spheroidal galaxy, and the fact that the distances to the majority of their member stars cannot be precisely measured, obtaining a three-dimensional profile is difficult. Previously, Sagittarius has had its line-of-sight width measured using the same OGLE-IV dataset as our analysis \citep{hamanowicz:2016AcA....66..197H}, but only the Magellanic Clouds have had their complete three-dimensional structure directly measured from samples of RR Lyrae stars~\citep{debsmc:2017arXiv170703130D,deblmc:2018MNRAS.478.2526D}. 

\par In this paper, we make the first measurement of the complete three-dimensional structure of the core of Sagittarius. We use two samples of RR Lyrae: one from the OGLE-IV bulge survey and one from \emph{Gaia} DR2. The OGLE data have precise three-dimensional positions, and so contain information on the three-dimensional stellar distribution in the core. The \emph{Gaia} sample of RR Lyrae do not have precise distances, but they form the most homogenous and spatially complete sample of RR Lyrae in Sagittarius. Therefore they contain information on the projection of Sagittarius onto the two-dimensional sky plane. 

\par To fit the data we model the core as a full triaxial ellipsoid. We constrain the scale lengths of the spheroid and the inclination using the three-dimensional information from the OGLE data. We project the spheroid onto the two-dimensional sky plane, and use the~\emph{Gaia} data to obtain independent constraints on the scale lengths. We will show that the strongest constraints on the scale lengths and the inclination of the core are obtained with a joint analysis of the OGLE and~\emph{Gaia} data. 

\par This paper is organized as follows. In section~\ref{sec:Data} we describe in detail the \emph{Gaia} DR2 and OGLE-IV data, and discuss the cuts that we implement to obtain our final RR Lyrae sample. In section~\ref{sec:Methods} we outline our formalism for measuring the shape of the stellar distribution, and describe the statistical methodology used to compare different models. In section~\ref{sec:results} we present the results of our analysis, and in section~\ref{sec:discussion} we end with conclusions and discussion. 

\section{Data}
\label{sec:Data}

\par In this section we describe the selection of our data samples from the \emph{Gaia} and the OGLE catalogs, and our determination of the distances to the stars and their associated errors. 
\subsection{Selection of data from \emph{Gaia} DR2 variable catalog}\label{sec:data2d}
 \begin{figure*}
    \centering
    \includegraphics[width=0.98\textwidth]{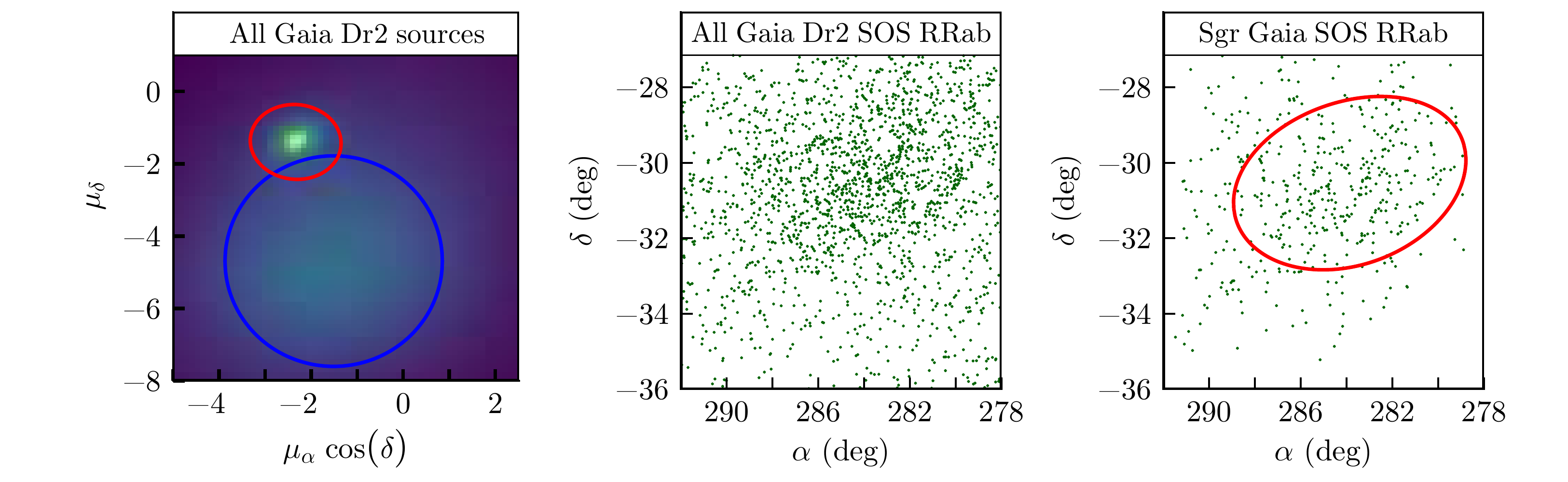}
    \caption{\emph{Left:} Density of all $8.5 \times 10^6$ Gaia sources in proper motion space. The solid red ellipse denotes a $1\, \sigma$ contour for the Sagittarius Gaussian component, and the solid blue line is the same for a Milky Way component.
    \emph{Middle:} The green points show all sources from the \emph{Gaia} DR2 variability catalog with a \emph{best\_classification} of RRab in the on sky region of Sagittarius. 
    \emph{Right:} The green points are the same as the left plot, but only the sources that pass our proper motion and magnitude cuts are plotted. The red ellipse shows the projected half light radius of Sagittarius found from RR Lyrae stars.}
    \label{fig:gaia_all}
\end{figure*}
 
 \par We use data from the second data release \citep{gaiaDR2:2018A&A...616A...1G} from the \emph{Gaia} mission \citep{Gaia_mission} to identify the 2D positions of a sample of RRab stars that are consistent with being members of Sagittarius. The \emph{Gaia} parallaxes are not measured well enough to obtain the distance precision required to constrain the 3D structure of Sagittarius. However, this dataset is much more spatially extended than the OGLE dataset. As we show below the larger spatial extent of these data allow us to obtain constraints on the three-dimensional properties that are complementary to the OGLE data.
 
\begin{figure}
    \centering
    \includegraphics[width=0.98\columnwidth]{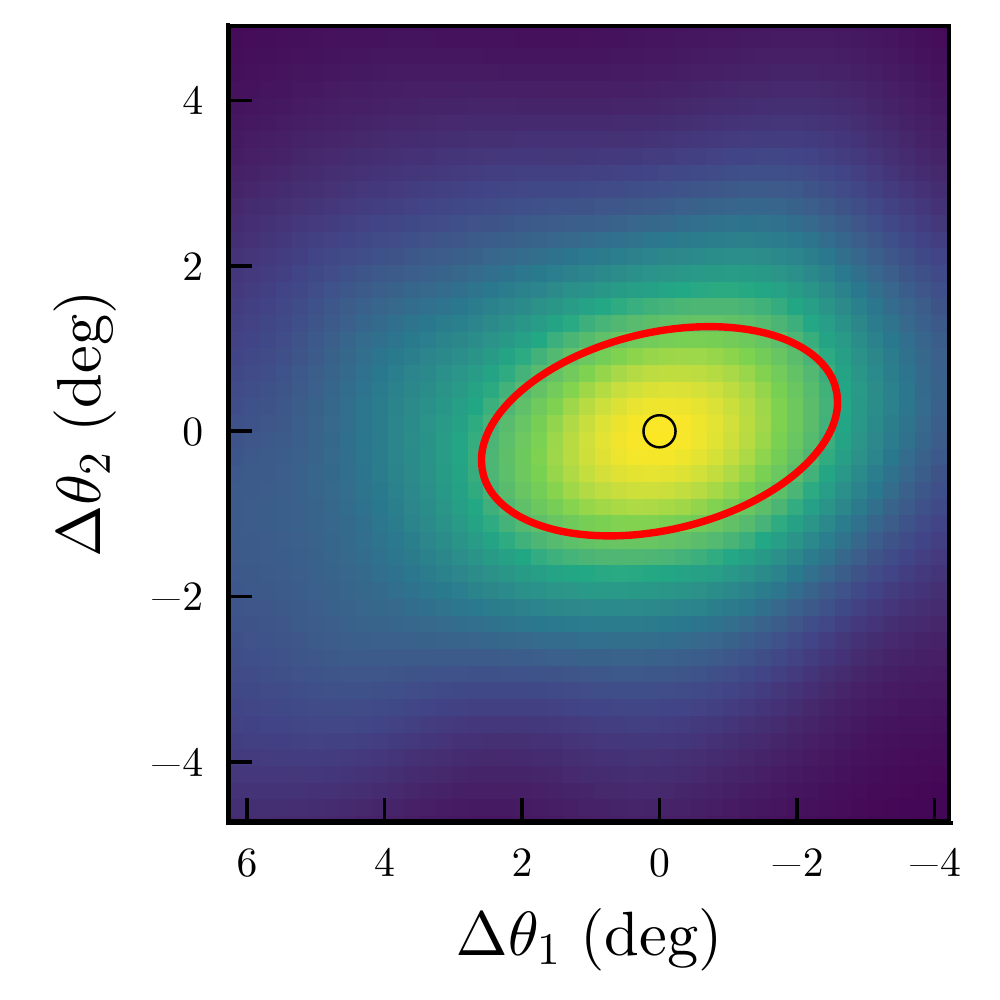}
    \caption{Shown is a kernel density estimate of \emph{Gaia} RR Lyrae (points in right plot of Figure \ref{fig:gaia_all}) as a function of angular position in the sky where yellow indicates a more dense region, and blue is less dense. $\Delta \theta_1$ and $\Delta \theta_2$ show the angular distance from the center of Sagittarius. The red ellipse indicates the projected half-light radius as Figure \ref{fig:gaia_all}, and the black circle marks the center of the galaxy.}
    \label{fig:kde_gaia}
\end{figure}{}

 \par To obtain a clean sample of \emph{Gaia} candidate RR Lyrae we start by using the full DR2 catalog to determine the appropriate proper motion cut to apply. We select all stars in \emph{Gaia} DR2 with a measured parallax ($\Bar{\omega}$) of  $<1\, \textrm{mas}$ that are within the core region of Sagittarius, defined by a right ascension ($\alpha$) and declination ($\delta$) of $279^\circ < \alpha < 291^\circ$ and $-40^\circ < \delta <-20^\circ$, respectively (See Appendix \ref{appendix:queries} for the query that we use). This selection criteria produce a sample of  $8.5 \times 10^6$ sources which we use to determine the proper motion of Sagittarius. The left panel of Figure \ref{fig:gaia_all} shows a 2D density histogram of the proper motions of these sources. A bimodal distribution can clearly be seen with a broad component due to the Milky Way and a much narrower one due to Sagittarius.
 
 \par  In order to determine the proper motion cut necessary to separate stars that are associated with Sagittarius from those that are likely associated with the Milky Way we fit a simple two Gaussian mixture model to the data. We define $\bs{\mu}$ as the centroid of the Sagittarius component, the components dispersion as $\bs{\sigma}=(\sigma_{\textrm{major}},\sigma_{\textrm{minor}})$, and a rotation by $\theta$, the angle between the $\mu_{\alpha}\cos(\delta)$ axis and the $\sigma_{\textrm{major}}$ axis. The left panel of Figure~\ref{fig:gaia_all} also shows the two-dimensional Gaussian model fit to the proper motion of these sources. The blue ellipse shows the $1\sigma$ best-fit component for the Milky Way population, and the red ellipse marks the $1\sigma$ best fit parameters for the Sagittarius component. For the Sagittarius component we find best-fit values of  $\bs{\mu}=(-2.34,-1.40)\,[\textrm{mas/yr}]$, $\sigma_{major}=0.52\,[\textrm{mas/yr}]$, $\sigma_{minor}=0.49\,[\textrm{mas/yr}]$, and $\theta=17.7\,[\deg]$. The mean proper motion of the Sagittarius component is consistent with the results of \citet{Helmi:2018A&A...616A..12G} and \citet{Fritz:2018A&A...619A.103F}. 
 We use these proper motion fit parameters below to identify RRab stars from the \emph{Gaia} sample that are members of Sagittarius.

 \par Next we query the \gaia{} Specific Objects Study (SOS) catalog \citep{ gaia_rrl:2019A&A...622A..60C} to obtain the sample of stars classified as RRab stars in the region of Sagittarius. Quality cuts are applied on the RRL catalog following \citet{Iorio:2019}.   This sample is shown in the middle panel of Figure~\ref{fig:gaia_all}. To select candidate RR Lyrae that are consistent with being Sagittarius members we apply a proper motion cut based on the fit to all sources in the area. Only stars with proper motions within $2\sigma$ of the Saggitarius mean proper motion are kept. We further remove all stars within the tidal radius of M54 as identified in \citet{hamanowicz:2016AcA....66..197H} ($\alpha=283.76^\circ,\;\delta=-30.48^\circ, \; r_{\textrm{tidal}}=7'.5$). Additionally, we apply a raw magnitude cut of $M_G > 16.8$ to remove any Milky Way foreground. This cleaned sample is shown in the right panel of Figure \ref{fig:gaia_all}.

 \par Finally, the positions of these stars are converted from ($\alpha,\delta$) to Sagittarius-centered coordinates ($\rho, \phi$), where $\rho$ is the angular separation between a star at ($\alpha,\delta$), the centroid of Sagittarius ($\alpha_0,\delta_0$), and $\phi$ is the position angle of a star with respect to the centroid. The points are then projected onto an angular plane similar to equations 1-4 in \citet{vandermarel:2001AJ....122.1807V}. Next, we perform a Gaussian kernel density estimate (KDE) with a $1\;\textrm{deg}^2$ window as a rough check on the distribution of these stars. This KDE in Figure \ref{fig:kde_gaia} shows a density spike at the location of Sagittarius that drops off in a manner consistent with being Gaussian. We apply a rectangular cut of $-4.2 < \Delta\theta_1[\textrm{deg}] < 6.2$ and $-4.7 < \Delta\theta_2[\textrm{deg}] < 4.9$ where only the 482 stars within this area are used in our analysis. 
 \begin{figure*}
     \centering
     \includegraphics[width=0.66\textwidth]{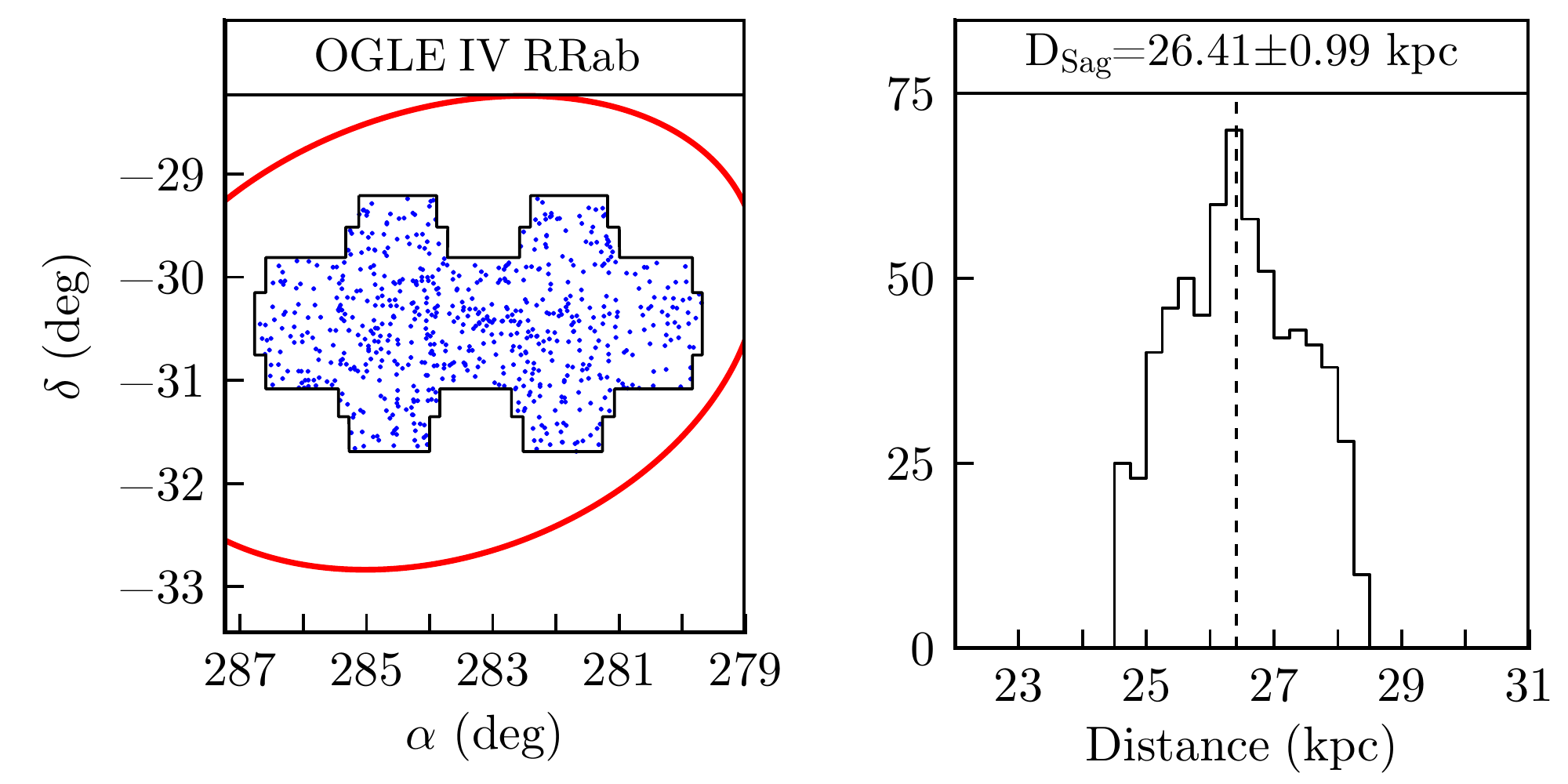}
     \caption{\emph{Left:} The blue points show the on-sky spatial distribution of the 670 OGLE RR Lyrae used in our analysis. The red ellipse shows the same projected half-light radius as Figure \ref{fig:gaia_all}, and the black outline shows the region where OGLE data was taken. \emph{Right:} A histogram of heliocentric distance for all stars in left-hand plot. The median distance of Sagittarius RR Lyrae is marked with a dashed line.}
     \label{fig:ogle_spatial_dist}
 \end{figure*}{}

\subsection{Selection of data from OGLE-IV RRab catalog}\label{sec:Data3d}

We use the OGLE-IV Bulge~\citep{ogleIV:2015AcA....65....1U} catalog of fundamental mode RR Lyrae variable stars (RRab, \citealt{ogleIVRRAB:2014AcA....64..177S,hamanowicz:2016AcA....66..197H}) for the 3D portion of our analysis. We clean the sample and derive distances to the RR Lyrae in the same manner as \citet{Jacyszyn-Dobrzeniecka:2017AcA....67....1J} and \citet{Skowron:2016AcA....66..269S}. Appendix~\ref{appendix:distance} provides a brief summary of the process; for more details we refer to \citet{Jacyszyn-Dobrzeniecka:2017AcA....67....1J} and references therein. 

\par We estimate the uncertainty on these distance measurements as follows. The statistical component is due to the accuracy of mean brightness in the $I$-band and $V$-band. From \citet{ogleIV:2015AcA....65....1U} the mean accuracy of these measurements is $\sigma_{I,V}=0.02$ mag, and the systematic uncertainty is introduced from the calculation of {\rm [Fe/H]} and the \citet{Braga:2015ApJ...799..165B} Period Luminosity Metallicity (PLZ) relation. 
Similar to \citet{Jacyszyn-Dobrzeniecka:2017AcA....67....1J} we take total uncertainty to be $3\%$ of the distance to a star.

\par After cleaning the catalog and deriving distances we make cuts to remove Milky Way foreground/background stars. First, the same proper motion cut discussed in the previous section is applied. Next, we select all RRab stars with $279.6^\circ < \alpha < 286.7^\circ$ and $-31.7^\circ < \delta < -29.23^\circ$ near the core of Sagittarius. Then, a cut is applied to remove stars with large radial distances from the core ($r > 2\; \textrm{kpc}$). This cut is implemented by defining $D_{\textrm{min}}$ ($D_{\textrm{max}}$) which corresponds to the minimum (maximum) heliocentric distance of a star with $r < 2\; \textrm{kpc}$ and removing all stars with distances outside of $D_{\textrm{min}} < D < D_{\textrm{max}}$. 

\par We define the center of Sagittarius to be $\alpha_0=283.83^\circ$, $\delta_0=-30.55^\circ$ \citep{hamanowicz:2016AcA....66..197H, mcconnachie:2012AJ....144....4M} and the distance to be the median distance of our sample $D_0=26.41\; \textrm{kpc}$. 
This distance differs from \citet{hamanowicz:2016AcA....66..197H} who found $D_0=26.98\;\textrm{kpc}$ due to the slightly different method of distance estimation to the RR Lyrae (we are using Wesenheit magnitudes instead of I-band magnitudes). Finally, member stars of the globular cluster M54 are removed in the same manner detailed above. The positions and distances of the 721  RR Lyrae stars that compose our clean sample are shown in Figure~\ref{fig:ogle_spatial_dist}.

\par Additionally, prior to our analysis we compare the \emph{Gaia} sample to the OGLE sample to check the \emph{Gaia} sample for completeness and contamination. From \citet{gaia_rrl:2019A&A...622A..60C} the \emph{Gaia} SOS catalog is 15\% complete with 9\% contamination in the OGLE bulge fields for RRab, RRc, and RRd stars. To get a sense of the completeness and contamination in our sub-sample, which is much fainter on average ($\mean{M_G}=18.14$) than the full OGLE bulge catalog, we cross-match the cleaned \emph{Gaia} SOS catalog with the less extended, but more complete and pure OGLE Sgr sample. There are 123 stars in common between the two datasets. We find that for this much smaller sample the \emph{Gaia} catalog is $15\%$ complete with $2\%$ contamination and importantly has no obvious spatial structure in the completeness or the contamination. Due to the low contamination rate and absence of artificial structure in the data, we assume the distribution is unbiased, and can be used to infer projected properties of the Sagittarius dwarf galaxy.

\section{Methods} \label{sec:Methods}
\par In this section we present a model for the 3D properties of the RRab distribution in the Sagittarius core. We then move on to define the likelihood functions that we use for inference with both the 2D \emph{Gaia} DR2 data and 3D OGLE data. 

\subsection{Modeling the 3D properties of Sagittarius} \label{sec:3danalysis}

\par We define an inertial coordinate system centered on Sagittarius such that $\hat{x}$ is aligned along the major axis, $\hat{y}$ along the intermediate axis, and $\hat{z}$ along the minor axis. In this coordinate system, we model the 3D stellar distribution of RR Lyrae as a Gaussian,
\begin{equation}
    \rho_{RRL}=\rho_0\, \exp\bigg(-\frac{1}{2}\bigg[\frac{x^2}{a^2}+\frac{y^2}{b^2}+\frac{z^2}{c^2}\bigg]\bigg), 
    \label{eq:Gaussian}
\end{equation}
where $a, b, c$ are the scale lengths of the density distribution in the respective coordinate directions, and $\rho_0$ is a scale density. The radius along the major axis for any location is then given by $r=\sqrt{{x^2}/{a^2}+{y^2}/{b^2}+{z^2}/{c^2}}$. For this three-dimensional Gaussian density distribution, the half-light radius along the major axis is given by $1.56\times a$. We choose the Gaussian form in Equation~\ref{eq:Gaussian} because its physical interpretation is straightforward, and also because its analytic properties can be readily determined in convolutions that we perform below. 

\par From the scale lengths ($a,b,c$), we can define the triaxility, $T$, in terms of the axis ratios $p = b/a$ and $q = c/a$
\begin{eqnarray}
T &=& \frac{1-p^2}{1-q^2}. 
\end{eqnarray}
For $T = 0$, the shape of the system is an oblate spheroid ($a=b > c$), while for $T=1$ the system is a prolate spheroid ($a > b = c$).

\par In order to connect to an observer-based coordinate system, we define a separate right-handed ``primed" coordinate system centered on Sagittarius. We convert the observed positions of the RR Lyrae to cylindrical coordinates in the plane of the sky, ($\rho,\phi$), as described in \citet{vandermarel:2001AJ....122.1807V}. We similarly define our coordinate system such that the $\hat{x}'$ axis points anti-parallel to the right ascension axis, the $\hat{y}'$ points parallel to the declination axis and the $\hat{z}'$ axis to points towards the observer on Earth. These transformations from ($\rho,\phi,D$) to ($x^\prime,y^\prime,z^\prime$) are given by 
\begin{align*}
x^\prime=&D\; \sin\;\rho \cos \phi\\
y^\prime=&D\; \sin\;\rho \sin \phi\\
z^\prime=&D_0-D\; \cos\;\rho
\end{align*}
with $D$ defined as the distance to a star and $D_0$ as the distance to the galaxy. 

To then transform from the inertial Sagittarius coordinate system to this observed frame we follow a similar formalism to \citet{Sanders2017MNRAS.472.2670S}. The transformation matrix between the coordinates is defined as the Euler rotation $\bs{R}(\alpha,\beta,\kappa)=\bs{R}_z(\alpha)\bs{R}_x(\beta)\bs{R}_z(\kappa)$, with the rotation matrices defined as 
\begin{align}\label{eqn:rotdef}
  \bs{R}_z(\omega)&=
  \begin{bmatrix}
  \cos(\omega) & -\sin(\omega)  & 0\\
  \sin(\omega) & \cos(\omega)   & 0\\
  0       & 0          & 1
  \end{bmatrix}
  ,\\
  \bs{R}_x(\omega)&=
  \begin{bmatrix}
  1 &  0      & 0\\
  0 & \cos(\omega) & \sin(\omega)\\
  0 & \sin(\omega) & \cos(\omega)
  \end{bmatrix}
\end{align}
For the inertial ($\bs{x}$) to observed ($\bs{x}^\prime$) transformation similar to~\citet{Xu:2019arXiv190408949X} we  define the observer to be located in the direction
\begin{equation*}
    \hat{z}^\prime \equiv \sin\theta\cos\phi\,\hat{x}+\sin\theta\sin\phi\,\hat{y}+\cos\theta\,\hat{z}
\end{equation*}
Then if ($x^\prime,y^\prime,z^\prime$) is the right handed coordinate system defined above the total transformation is $\bs{R}_{\textrm{int,obs}}=\bs{R}(\gamma,\pi/2-\phi,\theta)$ where $\gamma$ is a rotational degree of freedom in the plane of the sky. Written out the transformation is
\begin{align}\label{eqn:rotation}
\begin{bmatrix}
x^\prime\\
y^\prime\\
z^\prime\\
\end{bmatrix}
=&
\begin{bmatrix}
\cos\gamma & -\sin\gamma & 0 \\
\sin\gamma & \cos\gamma  & 0 \\
0          & 0           & 1
\end{bmatrix}
\times\nonumber\\
&\begin{bmatrix}
\sin \phi & -\cos\phi & 0 \\
\cos\theta\cos\phi & \cos\theta\sin\phi& -\sin\theta\\
\sin\theta\cos\phi & \sin\theta\sin\phi& \cos\theta
\end{bmatrix}{}
\begin{bmatrix}
x\\
y\\
z
\end{bmatrix}.
\end{align}

\par From the above definitions we must compute the projected properties of the galaxy in order to compare with the \emph{Gaia} sample. The formalism to obtain the projected positions is similar to \citet{Sanders2017MNRAS.472.2670S}; for our analysis we are particularly interested in expressing the result in terms of the projected major axis $a_{\textrm{proj}}$, the minor axis $b_{\textrm{proj}}$, the observed ellipticity $\epsilon=1-b_{\textrm{proj}}/a_{\textrm{proj}}$, and the position angle between observed North and the projected major axis $P.A.$. A derivation of the projected properties from the 3D model is found in Appendix \ref{appendix:projprop}.

\par Finally, we are interested in the inclination of Sagittarius with respect to both an Earth-based observer and one at the Galactic Center. To measure this we define the inclination angle ($i$) to be the dot product of the unit vector pointing along the major axis ($\bf{\hat{a}}$) and the vector that points from Sgr to the observer ($\bf{\hat{u}}$). Then the inclination is defined by $cos(i+\pi/2)={\bf\hat{a}}\boldsymbol{\cdot} \bf{\hat{u}}$ where $i=0$ indicates the major axis of Sgr is in the plane of the sky with respect to an observer. For an Earth-based observer in the ($\prime$) frame ${\bf\hat{u}}_{earth}=[0,0,1]$ and we take the distance between the sun and the Galactic Center to be $8.17\; kpc$ \citep{gravity:2019A&A...625L..10G} giving a ${\bf\hat{u}}_{GC}=[-0.11,0,0.99]$.

\subsection{Likelihood analysis}
\par With the model for Sagittarius outlined above, we now move on to discussing our likelihood analysis. Our likelihood analysis will involve a separate analysis of the 2D \emph{Gaia} and the 3D OGLE data, as well as a joint analysis of these data sets. 

\par We define the 2D and 3D data vectors as ${\cal D}_{2D}$ and ${\cal D}_{3D}$, respectively. We use these data sets to constrain our model parameters, which we take as the 3D parameters, $\bs{\Theta}_{3D}  =[{a,p,q,\gamma,\phi,\theta}]$, defined as above. We choose to use the 3D parameters as our base set of model parameters in order to analyze the 2D and 3D data in a consistent manner. The probability for the model parameters given the ${\cal D}_{3D}$ data is $P(\bs{\Theta}_{3D}\given {\cal D}_{3D})$, and similarly the probability for the model parameters given the ${\cal D}_{2D}$ data is $P(\bs{\Theta}_{3D}\given {\cal D}_{2D})$. 

Starting with the 2D case we need to derive an expression for $P(\bs{\Theta}_{3D}\given {\cal D}_{2D})$. In this case the data vector ${\cal D}_{2D}$ is given in the primed coordinates defined in the frame of the observer. We start by writing the observed Cartesian position of a star when projected into the $z^\prime=0$ plane as $\bs{x}^{\prime T}=[x^\prime,y^\prime,0]$.
Then given the Gaussian model for the core of Sagittarius the probability of a single star being observed at location $\bs{x}^{\prime}$ is 

\begin{equation}
    P(\bs{x}^\prime \given\bs{\Theta}_{3D})=\norm(\bs{x}^\prime \given\mean{\bs{x}^\prime}_{\textrm{Sgr}},\bs{C}_{\textrm{proj}}).
\end{equation}
where $\norm$ is defined as a multivariate normal distribution. In the above equation $\mean{\bs{x}^\prime}_{\textrm{Sgr}}$ is taken to be the center of Sagittarius as defined in Section \ref{sec:Data3d} which is $(0,0,0)$ in the prime ($\prime$) frame. Then $\bs{C}_{\textrm{proj}}$ is the covariance matrix
\begin{equation}
    \bs{C}_{\textrm{proj}}=\bs{R}_z^{T}(P.A.)
    \begin{bmatrix}
    a_{\textrm{proj}}^2  & 0             & 0\\
    0           & b_{\textrm{proj}}^2    & 0\\
    0           & 0             & 0
    \end{bmatrix}
    \bs{R}_z(P.A.).
\end{equation}
The covariance matrix is a function of the position angle ($P.A.$)  and the projected major axes $(a_{\textrm{proj}}^2, b_{\textrm{proj}}^2)$. These projected quantities can be derived from our model parameters as discussed in the previous section via the equations in Appendix \ref{appendix:projprop}.

In order to account for incomplete sampling of the stars in Sagittarius, we must develop a selection function to incorporate into the likelihood analysis. We define a simple selection function $\mathcal{S}_{2D}(x^\prime,y^\prime)=\int \mathcal{S}_{3D}(\bs{x}^\prime)\, \dd z$ that is equal to 1 inside the region within figure~\ref{fig:kde_gaia}, and 0 elsewhere. With this selection function, the probability of any set of $\bs{\Theta}_{3D}$ is then: 
\begin{align}
 P(\bs{\Theta}_{3D}| {\cal D}_{2D})
 &=\frac{\prod_{n}^{N}\mathcal{S}_{2D}(\bs{x^\prime})\;P(\bs{x^\prime}_n \given \bs{\Theta}_{3D})[\bs{\Theta}_{3D}]}
 {\int\int \mathcal{S}_{2D}(\bs{x^\prime}) P(\bs{x^\prime} \given \bs{\Theta}_{3D})\,\dd x^\prime \dd y^\prime}.
 \label{eq:2Dlikelihood}
\end{align}
where $N$ is the number of stars in the \emph{Gaia} sample and $n$ labels an individual star. With the likelihood defined above, we are able to conduct parameter inference on the 3D model parameters using the methods described below. 

\par Our 3D analysis that uses the OGLE data follows a similar formalism as in the 2D case. We define the true position of a single star as $\truepos^\prime$, and the observed position as $\bs{x}^\prime$. We assume that the true position of a star given its observed position follows the distribution, 

\begin{equation}
\label{eqn:obspos}
    P(\bs{x}^\prime\given\truepos^\prime)=\norm(\bs{x}^\prime\given\truepos^\prime,\bs{C}_{\bs{x}}). 
\end{equation}
Here $\bs{C}_{\bs{x}}$ is empirically determined by Monte Carlo resampling the distance of an individual star and recomputing its Cartesian position 5000 times, using the  assumed errors on the heliocentric distance of 3\% (see Section~\ref{sec:Data}) and taking into account the correlations between coordinates.

\par The population as a whole we model as a multivariate Gaussian with mean given by the centroid of the galaxy ($\mean{\bs{x}^\prime}_{\textrm{Sgr}}$), a dispersion defined by the three principal axes, and an axis of symmetry parameters p and q. 
The covariance matrix in the inertial frame is then: 
\begin{equation}
\bs{C}_{\textrm{Sgr}}=
\begin{bmatrix}
a^2 & 0 & 0\\
0 & b^2 & 0\\
0 & 0 & c^2
\end{bmatrix}
=
\begin{bmatrix}
a^2 & 0 & 0\\
0 & (p\times a)^2 & 0\\
0 & 0 & (q\times a)^2
\end{bmatrix}.
\end{equation}
Therefore, the probability of a star existing at true location $\truepos^\prime$ is given by:
\begin{equation}
\label{eqn:truepos}
    P(\truepos^\prime\given \bs{\Theta}_{3D})=\norm(\truepos^\prime\given\mean{\bs{x^\prime}}_{\textrm{Sgr}},\textbf{R}^{-1}\bs{C}_{\textrm{Sgr}}\textbf{R}). 
\end{equation}
Since both the observed (Equation \ref{eqn:obspos}) and true (Equation \ref{eqn:truepos}) components are Gaussian we can analytically marginalize the true positions of the stars out. Then the probability of observing a star at position $\bs{x}$ conditional on our model parameters is
\begin{equation}
\begin{aligned}
    P(\bs{x}^\prime \given \bs{\Theta}_{3D} )=&P(\bs{x}^\prime\given\truepos^\prime)\,P(\truepos^\prime\given \bs{\Theta}_{3D})\\
    =&\norm(\bs{x}^\prime\given\mean{\bs{x}^\prime}_{\textrm{Sgr}},\bs{C}_{\bs{x}^\prime_j}+\textbf{R}^{-1}\bs{C}_{\textrm{Sgr}}\textbf{R}).
\end{aligned}
\end{equation}
Similar to the 2D case, we then define a selection function as $\mathcal{S}_{3D}(\bs{x})$. In the 3D case, the selection function is 1 inside the area within the black outline shown in Figure 3 and within a 2 kpc distance from the center of Sagittarius. The full likelihood with the observed selection function is then

\begin{equation}
    P(\bs{\Theta}_{3D}|{\cal D}_{3D})
    =\frac{\prod_{n}^{N}\mathcal{S}_{3D}(\bs{x}^\prime)P(\{\bs{x}^\prime\}_N \given \bs{\Theta}_{3D})[\bs{\Theta}_{3D}]}
    {\int\int\int \mathcal{S}_{3D}(\bs{x}^\prime)P(\bs{x}^\prime \given \bs{\Theta}_{3D})\dd x^\prime\dd y^\prime\dd z^\prime}. 
    \label{eq:3Dlikelihood}
\end{equation}

\begin{table}
    \begin{center}
    \begin{tabular}{ c | c }
        \hline
        Parameter & Prior \\
        \hline
        $a$(Major axis) & $U(0.1,10)\;\textrm{kpc}$ \\
        $p$ (b/a) & $U(0.1,1)$ \\
        $q$ (c/a)& $U(0.1,1 \times p$) \\
        $\gamma$ & $U(-90, 90)\; \textrm{deg}$ \\
        $\phi$  & $U(-90, 90)\; \textrm{deg}$ \\
        $\theta$ & $U(-90, 90)\; \textrm{deg}$\\
        \hline
    \end{tabular}
    \caption{Assumed priors for our set of baseline model parameters. Parameters are defined in Section~\ref{sec:Methods}.
    \label{tab:priors}}
    \end{center}
\end{table}

The discussion above is in the context of a separate analysis for both the 2D and the 3D likelihoods. We will also consider a joint analysis, in which we fit to the combined 2D and 3D data sets. Assuming that the data sets are independent, which is a reasonable assumption for the \emph{Gaia} and OGLE data, the joint probability for the model parameters given the data is 
\begin{equation}
P(\Theta_{3D}\given {\cal D}_{2D},{\cal D}_{3D})=P(\Theta_{3D}\given {\cal D}_{2D})P(\Theta_{3D}\given {\cal D}_{3D}).
\end{equation}

For both the separate and the joint likelihood analyses, we determine posterior probabilities for the parameters $\bs{\Theta}_{3D}$. To determine these posteriors, we use the nested sampler \texttt{PyMultinest} \citep{Buchner:2014A&A...564A.125B,Feroz:2008,Feroz:2009} with 500 live points to generate samples of the posterior. For each parameter the prior used in the fit is listed in table \ref{tab:priors}. Note that for $p$ and $q$, the lower bound on the prior is $0.1$.

\begin{table*} 
    \centering   
    \begin{tabular}{c|c|c|c|c|}
        \hline
        Model Parameters      &  joint fit                           & \emph{Gaia} only fit               & OGLE only 2d fit & OGLE only 3d fit \\
        \hline
        \hline
        $a \,[\textrm{kpc}]$    &  $ 1.76 _{ -0.12 }^{+ 0.15 } $   &  $ 2.01 _{ -0.19 }^{+ 0.33 } $   &  $ 1.71 _{ -0.37 }^{+ 1.37 } $   &  $ 1.36 _{ -0.16 }^{+ 0.99 } $ \\ \\
        $p $    &  $ 0.74 _{ -0.1 }^{+ 0.13 } $   &  $ 0.46 _{ -0.06 }^{+ 0.08 } $   &  $ 0.74 _{ -0.27 }^{+ 0.23 } $   &  $ 0.88 _{ -0.36 }^{+ 0.14 } $ \\ \\
        $q $    &  $ 0.43 _{ -0.06 }^{+ 0.05 } $   &  $ 0.31 _{ -0.14 }^{+ 0.1 } $   &  $ 0.42 _{ -0.23 }^{+ 0.24 } $   &  $ 0.64 _{ -0.25 }^{+ 0.2 } $ \\ \\
        $\gamma\,[\textrm{deg}]$    &  $ -11.51 _{ -3.54 }^{+ 3.87 } $   &  $ -9.35 _{ -40.39 }^{+ 52.43 } $   &  $ -1.67 _{ -54.27 }^{+ 57.2 } $   &  $ 0.35 _{ -35.18 }^{+ 31.49 } $ \\ \\
        $\phi\,[\textrm{deg}]$    &  $ -21.24 _{ -4.76 }^{+ 7.83 } $   &  $ 2.67 _{ -67.75 }^{+ 64.16 } $   &  $ 3.00_{ -58.24 }^{+ 56.66 } $   &  $ 10.67 _{ -47.74 }^{+ 40.45 } $ \\ \\
        $\theta\,[\textrm{deg}]$    &  $ 3.15 _{ -17.32 }^{+ 21.51 } $   &  $ -4.62 _{ -50.69 }^{+ 63.82 } $   &  $ -5.28 _{ -62.63 }^{+ 69.18 } $   &  $ 114.50 _{ -61.38 }^{+ 37.44 } $ \\ \\
        \hline
        Derived Parameters\\
        \hline
        $a_{\textrm{proj}}\,[\textrm{kpc}]$    &  $ 1.73 _{ -0.11 }^{+ 0.14 } $   &  $ 1.88 _{ -0.14 }^{+ 0.16 } $   &  $ 1.53 _{ -0.25 }^{+ 0.87 } $   &  $ 1.31 _{ -0.14 }^{+ 0.88 } $ \\ \\
        $b_{\textrm{proj}}\,[\textrm{kpc}]$    &  $ 0.86 _{ -0.03 }^{+ 0.03 } $   &  $ 0.86 _{ -0.03 }^{+ 0.03 } $   &  $ 1.15 _{ -0.29 }^{+ 0.22 } $   &  $ 1.07 _{ -0.25 }^{+ 0.16 } $ \\ \\
        $P.A.\,[\textrm{deg}]$    &  $ 102.19 _{ -2.46 }^{+ 2.44 } $   &  $ 101.90 _{ -2.1 }^{+ 2.14 } $   &  $ 98.49 _{ -76.19 }^{+ 63.95 } $   &  $ 109.91 _{ -91.92 }^{+ 58.2 } $ \\ \\
        $T$    &  $ 0.56 _{ -0.26 }^{+ 0.18 } $   &  $ 0.87 _{ -0.1 }^{+ 0.08 } $   &  $ 0.61 _{ -0.49 }^{+ 0.31 } $   &  $ 0.47 _{ -0.57 }^{+ 0.48 } $ \\ \\
        $r_{1/2} \,[\textrm{kpc}]$    &  $ 2.74 _{ -0.18 }^{+ 0.23 } $   &  $ 3.14 _{ -0.3 }^{+ 0.51 } $   &  $ 2.67 _{ -0.58 }^{+ 2.14 } $   &  $ 2.12 _{ -0.26 }^{+ 1.55 } $ \\ \\
        $r_{1/2,\,\textrm{proj}} \,[\textrm{kpc}]$    &  $ 2.04 _{ -0.13 }^{+ 0.16 } $   &  $ 2.22 _{ -0.16 }^{+ 0.18 } $   &  $ 1.80_{ -0.29 }^{+ 1.03 } $   &  $ 1.55 _{ -0.16 }^{+ 1.04 } $ \\ \\
       \hline
    \end{tabular}
    \caption{Results of our fits for the joint case as well as the \emph{Gaia} only, OGLE only 2d ($\alpha,\,\delta$) and OGLE only 3d ($\alpha,\,\delta,\, D_\odot$) cases. The values are the median for each parameter,
and the errors show the 68\% confidence interval. Histograms showing the posteriors for a selection of these parameters is shown in Figure \ref{fig:majaxcomp}. For the joint fit the $r_{1/2,\,\textrm{proj}}= 2.04\, \textrm{kpc}$ corresponds to an on sky major axis of $269\pm21'$.}
    \label{tab:triax_result}
\end{table*}{}

\begin{figure*}
    \centering
    \includegraphics[width=0.98\textwidth]{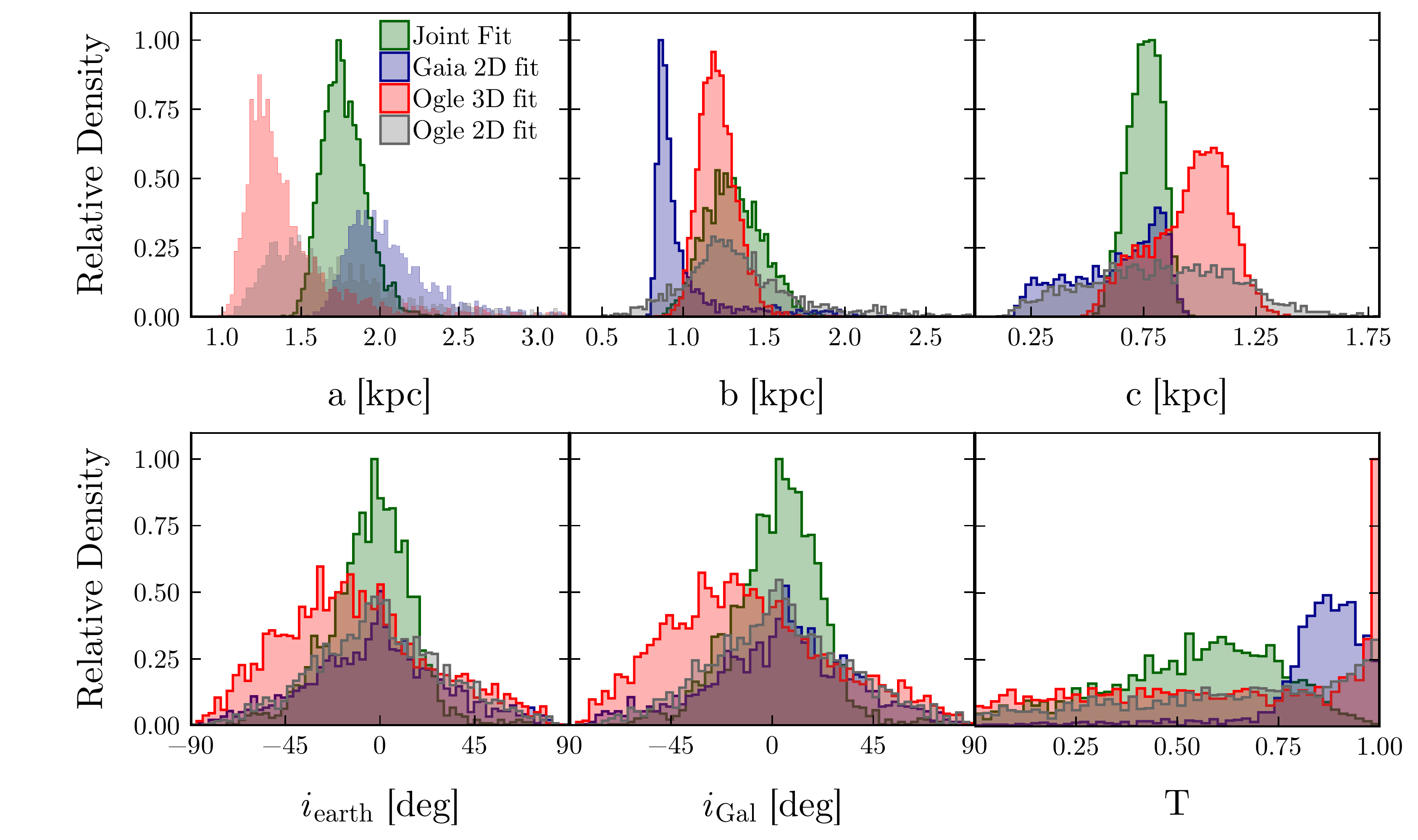}
    \caption{Posterior probability densities resulting from the likelihood analysis. The top row shows the results for the principal axes ($a,b,c$) of the core of Sagittarius, and the bottom row shows the posteriors for the inclination with respect to earth ($i_{\textrm{earth}}$) and with respect to the Galactic center ($i_{\textrm{Gal}}$ as well as the Triaxiality (T) of the system. To illustrate how each of the datasets effects our results posteriors are shown for the individual fits of the \emph{Gaia} (\emph{blue}) and OGLE data (\emph{red}), a fit of only the 2D information from the OGLE data (\emph{grey}), and the joint fit of the two datasets (\emph{green}). 
    \label{fig:majaxcomp}}
\end{figure*}{}
\section{Results}
\label{sec:results}
\par Figure~\ref{fig:majaxcomp} shows the posterior probability densities for the principal axes ($a,b,c$), inclination ($i_{\textrm{earth}}$), Galactic inclination ($i_{\textrm{Gal}}$) and triaxiality (T). The main result from a joint fit to the \emph{Gaia} and OGLE  data is shown as a green histogram, and a corner plot of this fit with the six model parameters is shown in appendix \ref{appendix:corner}. The median of the cumulative distributions and the 68\% containment intervals for each of the parameters are shown in Table~\ref{tab:triax_result}. We generally find that the scale parameters $a, p, q$ are well determined in the joint analysis, with the minor-to-major axis ratio $q$ being the best determined parameter, which is measured to $\lesssim 20\%$. The corresponding angles are generally less well constrained, in particular the rotation angle $\theta$ is not well determined by our analysis.

\begin{figure}
    \centering
    \includegraphics[width=0.98\columnwidth]{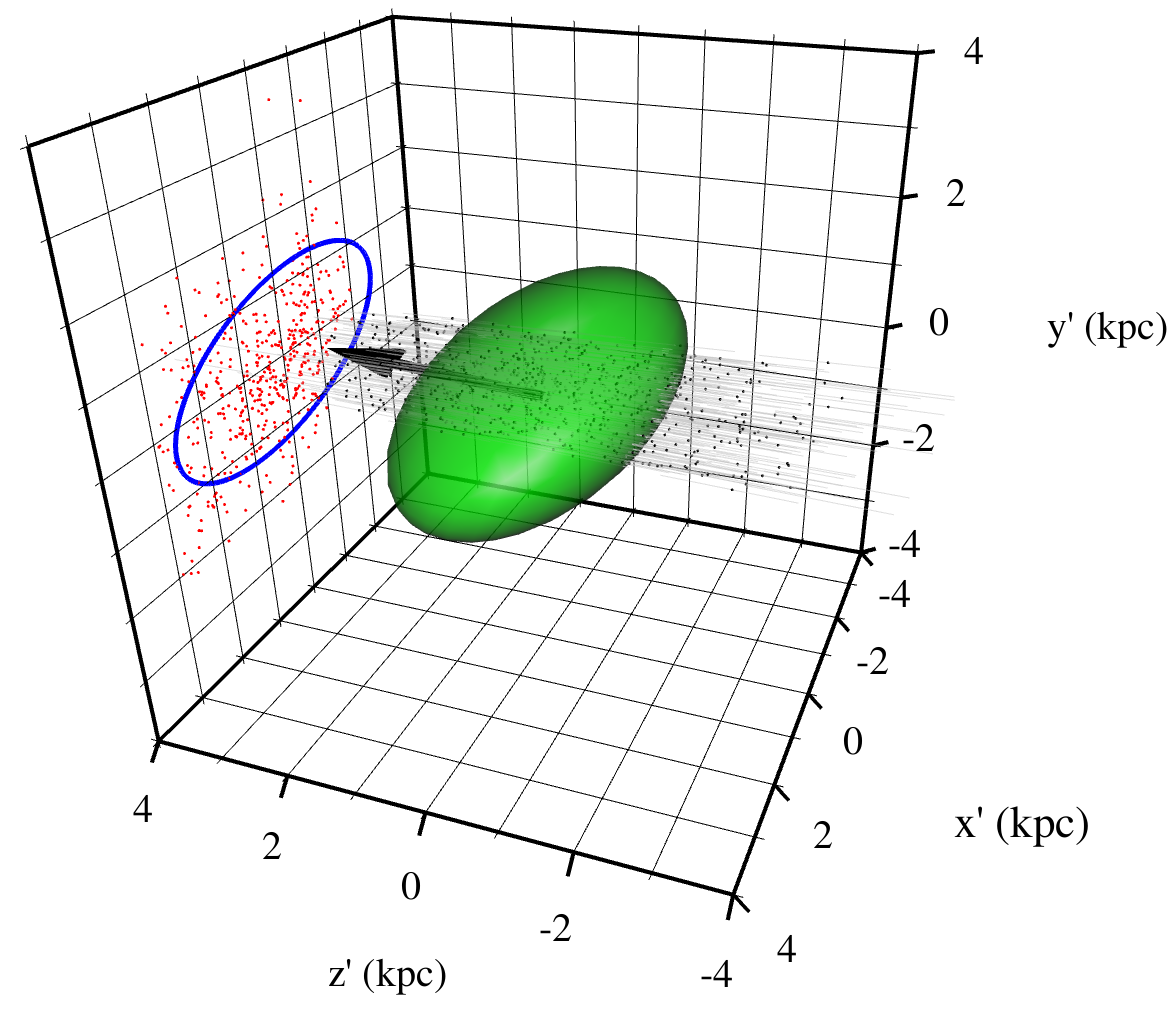}
    \caption{Three dimensional plot showing the results of our analysis. The red points show the two dimensional location of \emph{Gaia} RR Lyrae, and the blue ellipse indicates the half light radius derived from them. The black points show OGLE RR Lyrae where the grey lines indicate the uncertainty on the position of each star. The green ellipsoid shows the three dimensional half light radius of Sagittarius derived from the results of our analysis. The $\hat{z^\prime}$ axis points towards the observer. The black arrow points towards the Galactic center.}
    \label{fig:3dplots}
\end{figure}{}
\par To get a sense of which data set is providing more statistical constraining power, Table~\ref{tab:triax_result} shows results from the individual fits to each of the \emph{Gaia} data, the full 3D OGLE data, and only the 2D OGLE data. The medians of the scale lengths from the individual fits are found to differ; in particular the \emph{Gaia} data favors a major axis radius $\sim 2.01$ kpc, while the 3D OGLE data favors a lower value, $\sim 1.36$ kpc. However the results are consistent when considering the 68\% containment intervals. Figure~\ref{fig:majaxcomp} shows a comparison of the full posterior distributions of the major axis scale length as determined for both the joint and separate analyses. From this figure we see that the major axis scale length determined from the \emph{Gaia} data is more aligned with the results from the joint fit. This is because the \emph{Gaia} data is more extended, and thereby providing more constraining power than the OGLE data on the scale length.  

\par In Table~\ref{tab:triax_result} we also show measurements of parameters derived from our baseline set of parameters. The 3D half-light radii along the major axis is $r_{1/2}=1.17_{-0.06}^{+0.07}\,\textrm{kpc}$, and the 2D half-light radii along the major axis is $r_{1/2,\,proj}=1.14_{-0.06}^{+0.07}\,\textrm{kpc}$. Note that our fits are different than the half-light radius obtained from the red giant fits with 2MASS data~\citep{Majewski:2003ApJ...599.1082M}. The projected major axis of our fit is found to have a half light radius of $269\pm21'$ compared to their value of $342\pm12'$, and the projected ellipticity from our fit is $\epsilon = 0.49 \pm 0.04$, which can be compared to their result of $\epsilon = 0.62 \pm 0.02$. We do find consistent results for the position angle ($P.A.$).

\par From the parameters $p$ and $q$, we derive the posterior probability distribution of $T$. We find $T = 0.56 _{ -0.26 }^{+ 0.18 }$, where again the uncertainties are 68\% containment confidence intervals. Interestingly, a prolate spheroid corresponding to $T=1$ is ruled out by the model, while an oblate spheroid is strongly disfavored at the $\sim 95\%$ confidence level.

\par From the posterior probabilities we also deduce both the inclination of the major axis with respect to the plane of the sky, and the projection of the major axis on the direction towards the Galactic center. The inclination relative to the observer is $i_{earth}=-4.9_{-18.8}^{+17.5}\, \mathrm{deg}$, and the inclination relative to the Galactic center is $i_{Gal}=1.6_{-18.9}^{+17.5}\, \textrm{deg}$. Note again that an inclination of zero implies that the major axis is fully within the plane of the sky relative to the observer. These two values of the inclination are very similar due to the location of Sagittarius relative to the Galactic center. We note that the major axis is not aligned with the direction of the Galactic center, which may have important implications when comparing to general theoretical predictions for dwarf galaxies, which we discuss in more detail below. 

Figure~\ref{fig:3dplots} shows a three-dimensional view of the system. The red points are \emph{Gaia} RR Lyrae projected onto to the z=0 plane, and the blue ellipse is the projected half light ellipse. The black points show the 3D distribution of OGLE RR Lyrae in Sagittarius-centered Cartesian coordinates ($\bs{x}^\prime$), and the grey lines show the uncertainty on position for each OGLE star. The green ellipsoid marks the three dimensional half light radius as inferred by our analysis; its orientation and triaxiality can be seen. As a reminder the $\hat{z^\prime}$ axis points towards the observer. 

\begin{figure*}
    \includegraphics[width=0.98\textwidth]{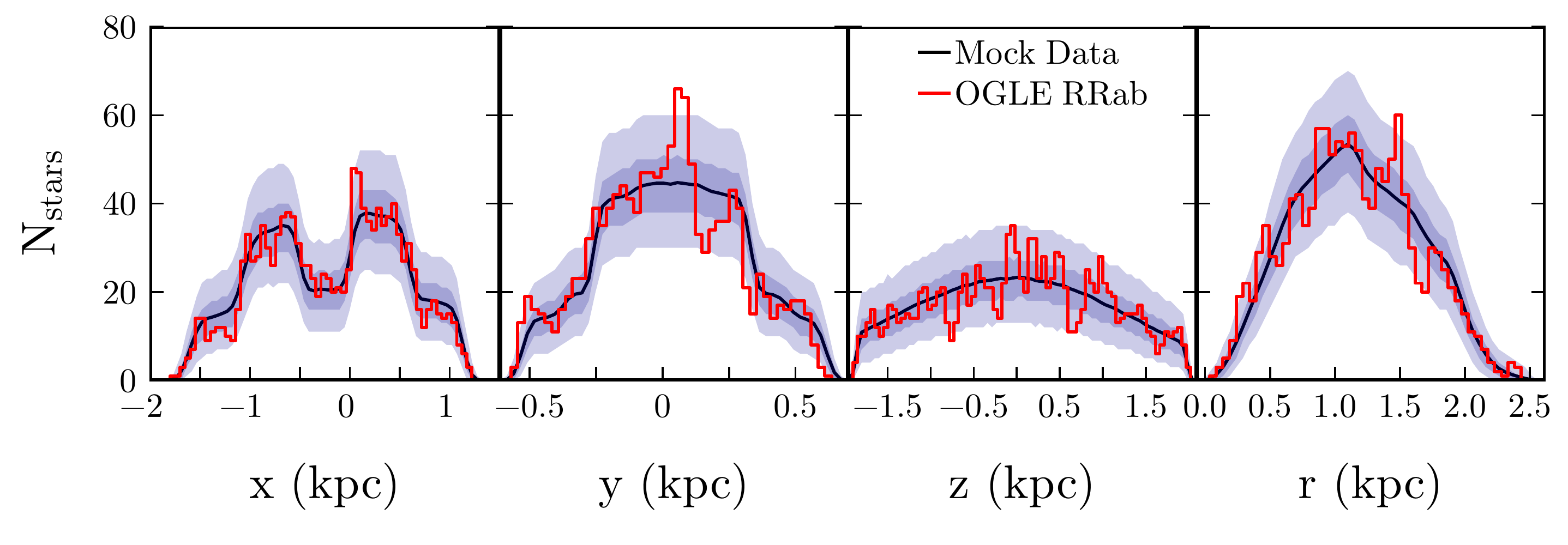}
    \includegraphics[width=0.55\textwidth]{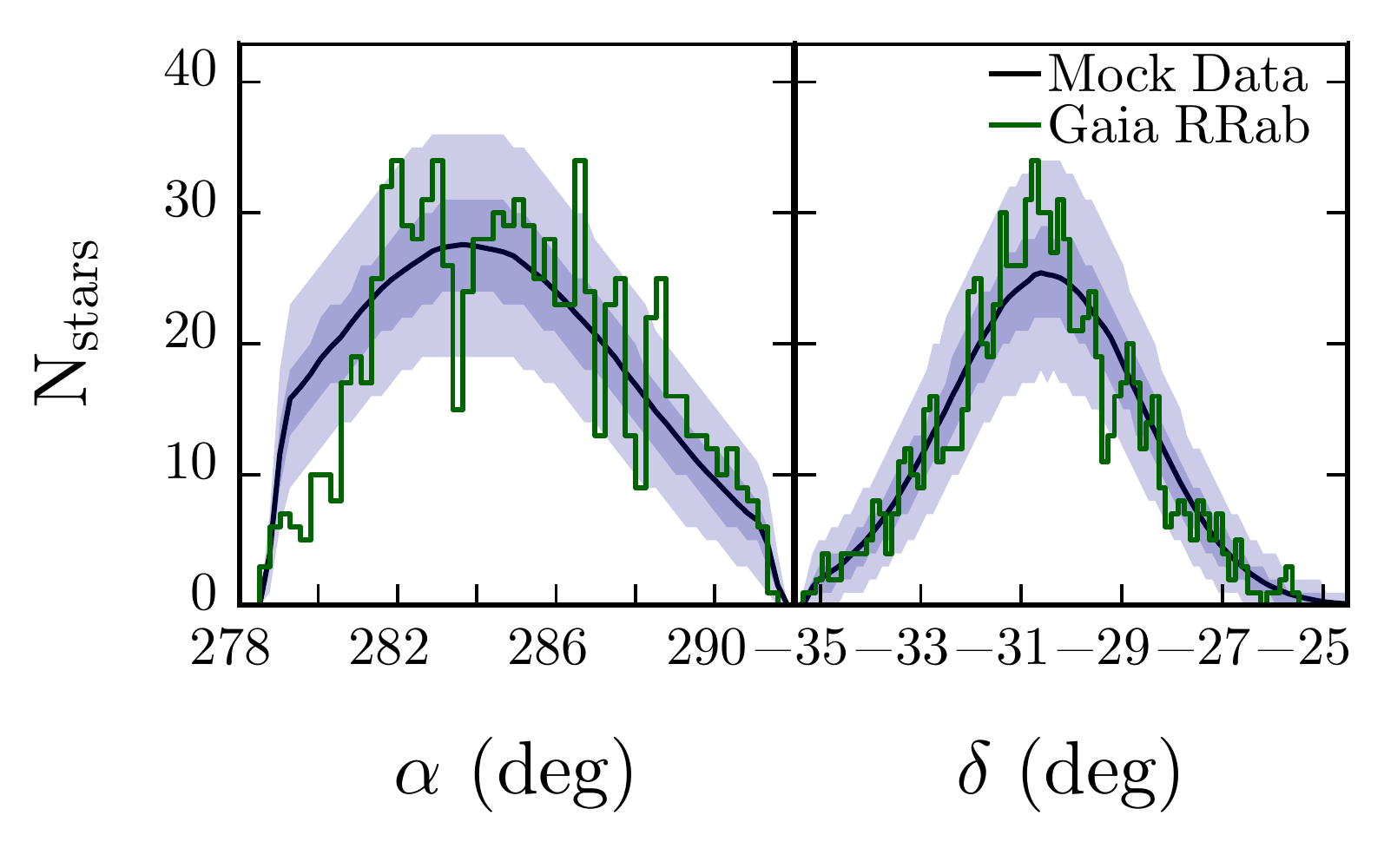}
    \caption{
    Histograms showing the distribution of OGLE RR Lyrae in different dimensions.
    For each panel the red (green) line shows the observed distribution of OGLE (\emph{Gaia}) stars, the black line is the median of 1000 mock realizations drawn from the posteriors of our model, and the blue contours show 50\% and 95\% confidence intervals for the mock realizations.
    \emph{Top row:} the panels show comparisons in x', y', z', and the radial distribution between the observed 3D positions of OGLE RR Lyrae and the model.\emph{Bottom row:} the panels compare the observed on sky ($\alpha,\,\delta$) positions of \emph{Gaia} RR Lyrae and the model.}
    \label{fig:mock_compare}
\end{figure*}

As a consistency check we take our posterior distribution of parameters $\Theta_{3D}$ and generate mock observed RR Lyrae distributions. This is done by sampling $\Theta_{3D}$ values randomly from the equal weighted posteriors (histograms in figure \ref{fig:jointcorner}), generating data following our model distribution, and applying the observational selection function. These mock catalogs should have similar spatial distributions as our observed data. The results of repeating this process for 1000 mock data sets are shown in Figure \ref{fig:mock_compare}, where in the top row the red line shows the observed distribution of OGLE RR Lyrae in Cartesian coordinates as well as the radial distribution of these stars stars. In the bottom row the green line shows the observed distribution of \emph{Gaia} RR Lyrae in both right ascension ($\alpha$) and declination ($\delta$). For all of these plots the black line marks the median distribution of our mock catalogs over the same coordinates, and the blue shading indicates 50\% and 95\% confidence intervals for these mock catalogs. We find that there is generally good agreement between the observed distribution and the expected distribution based on the results of our analysis. 
{}
\section{Discussion and Conclusion}
\label{sec:discussion}
\par We have performed the first 3D modeling of the spatial distribution of stars in the core of the Sagittarius dSph. Our sample of stars comes from both~\emph{Gaia} DR2 data and from the OGLE-IV RR Lyrae catalog. We derive distances to the OGLE stars, and from these and the \emph{Gaia} data we find that the spatial distribution is a triaxial ellipsoid. The more simple case of a prolate spheroid is ruled out at high statistical significance. These results come from a combined analysis of the~\emph{Gaia} and OGLE data-- such strong results are not attainable from an individual dataset. 

\par Our results may be compared to previous estimates of the three-dimensional structure of Sagittarius.~\citet{1997AJ....113..634I} use red clump stars and  find that the core is consistent with a prolate spheroid. The OGLE collaboration~\citep{hamanowicz:2016AcA....66..197H} used their RR Lyrae catalog~\citep{ogleIVRRAB:2014AcA....64..177S} to measure FWHM along the line-of-sight and find the size in this dimension to be 2.42 kpc. 

\par We have obtained the first measurement of the orientation of the stellar distribution of Sagittarius with respect to the plane of the sky and with respect to the Galactocentric frame of reference. The major axis of the RR Lyrae distribution is aligned nearly parallel to the sky plane, and the major axis is nearly perpendicular to the direction of the Galactic center. 

\par It is interesting to compare this result to the predictions of cosmological simulations~\citep{2007ApJ...671.1135K,2015MNRAS.447.1112B}. These generally find that the major axis of the dark matter distribution of subhalos is aligned with the Galactic center. This alignment is found to be stronger for systems that are closer to the Galactic center and for those that have been heavily tidally disrupted, i.e. systems like Sagittarius. These results from simulations however only apply to the dark matter distribution. It will be important in the future to develop predictions for the orientation of the dark matter distribution relative to the stellar distribution, and compare to the results we have obtained for Sagittarius. 

\par It is also interesting to compare our results to theoretical models for the Sagittarius progenitor. For example the elongated shape may be produced in models where the progenitor was a disk galaxy, and the system is currently at its second pericenter passage, transforming from a disk galaxy to a more spheroidal structure resembling a dwarf spheroidal~\citep{2010ApJ...725.1516L}. 

\par Since our RR Lyrae sample is mostly contained within the half-light radius of the Sagittarius core, the sample may be used to probe the dynamical mass distribution in the remnant core of the system that remains bound and in dynamical equilibrium. Three-dimensional positions combined with line-of-sight velocities of stars are expected to improve measurements of the velocity anisotropy parameter~\citep{Richardson:2013hga}. Targeted radial velocity measurements, along with  future improvements in \emph{Gaia} proper motions, would provide the first full six-dimensional phase space coverage of stars in a dwarf spheroidal, allowing for an unprecedented analysis of the dynamical state of the dark and luminous mass in the galaxy.

\section*{Acknowledgements}
We are grateful to Jennifer Marshall and Andrew Pace for discussions that improved this paper. We gratefully acknowledge support from the College of Science at TAMU through a Strategic Transformative Research Program (STRP) Seed Grant. The authors acknowledge the Texas A\&M University Brazos HPC cluster that contributed to the research reported here. This work has made use of data from the European Space Agency (ESA) mission
{\it Gaia} (\url{https://www.cosmos.esa.int/gaia}), processed by the {\it Gaia}
Data Processing and Analysis Consortium (DPAC,
\url{https://www.cosmos.esa.int/web/gaia/dpac/consortium}). Funding for the DPAC
has been provided by national institutions, in particular the institutions
participating in the {\it Gaia} Multilateral Agreement. This research made use of the PYTHON and R Programming Languages, along with many community-developed or maintained software packages including ASTROPY \citep{astropy:2013,astropy:2018}, NUMPY \citep{numpy},SCIPY \citep{2020SciPy-NMeth}, PYMULTINEST \citep{Buchner:2014A&A...564A.125B,Feroz:2008,Feroz:2009} and GGPLOT2 \citep{ggplot}.

\bibliographystyle{mnras}
\bibliography{sag_rrl}

\begin{thebibliography}{}
\makeatletter
\relax
\def\mn@urlcharsother{\let\do\@makeother \do\$\do\&\do\#\do\^\do\_\do\%\do\~}
\def\mn@doi{\begingroup\mn@urlcharsother \@ifnextchar [ {\mn@doi@}
  {\mn@doi@[]}}
\def\mn@doi@[#1]#2{\def\@tempa{#1}\ifx\@tempa\@empty \href
  {http://dx.doi.org/#2} {doi:#2}\else \href {http://dx.doi.org/#2} {#1}\fi
  \endgroup}
\def\mn@eprint#1#2{\mn@eprint@#1:#2::\@nil}
\def\mn@eprint@arXiv#1{\href {http://arxiv.org/abs/#1} {{\tt arXiv:#1}}}
\def\mn@eprint@dblp#1{\href {http://dblp.uni-trier.de/rec/bibtex/#1.xml}
  {dblp:#1}}
\def\mn@eprint@#1:#2:#3:#4\@nil{\def\@tempa {#1}\def\@tempb {#2}\def\@tempc
  {#3}\ifx \@tempc \@empty \let \@tempc \@tempb \let \@tempb \@tempa \fi \ifx
  \@tempb \@empty \def\@tempb {arXiv}\fi \@ifundefined
  {mn@eprint@\@tempb}{\@tempb:\@tempc}{\expandafter \expandafter \csname
  mn@eprint@\@tempb\endcsname \expandafter{\@tempc}}}

\bibitem[\protect\citeauthoryear{{Astropy Collaboration} et~al.,}{{Astropy
  Collaboration} et~al.}{2013}]{astropy:2013}
{Astropy Collaboration} et~al., 2013, \mn@doi [\aap]
  {10.1051/0004-6361/201322068}, \href
  {http://adsabs.harvard.edu/abs/2013A%26A...558A..33A} {558, A33}

\bibitem[\protect\citeauthoryear{{Barber}, {Starkenburg}, {Navarro}  \&
  {McConnachie}}{{Barber} et~al.}{2015}]{2015MNRAS.447.1112B}
{Barber} C.,  {Starkenburg} E.,  {Navarro} J.~F.,   {McConnachie} A.~W.,  2015,
  \mn@doi [\mnras] {10.1093/mnras/stu2494}, \href
  {https://ui.adsabs.harvard.edu/abs/2015MNRAS.447.1112B} {447, 1112}

\bibitem[\protect\citeauthoryear{{Battaglia}, {Helmi}  \&
  {Breddels}}{{Battaglia} et~al.}{2013}]{2013NewAR..57...52B}
{Battaglia} G.,  {Helmi} A.,   {Breddels} M.,  2013, \mn@doi [\nar]
  {10.1016/j.newar.2013.05.003}, \href
  {https://ui.adsabs.harvard.edu/abs/2013NewAR..57...52B} {57, 52}

\bibitem[\protect\citeauthoryear{{Belokurov} et~al.,}{{Belokurov}
  et~al.}{2014}]{2014MNRAS.437..116B}
{Belokurov} V.,  et~al., 2014, \mn@doi [\mnras] {10.1093/mnras/stt1862}, \href
  {https://ui.adsabs.harvard.edu/abs/2014MNRAS.437..116B} {437, 116}

\bibitem[\protect\citeauthoryear{{Braga} et~al.,}{{Braga}
  et~al.}{2015}]{Braga:2015ApJ...799..165B}
{Braga} V.~F.,  et~al., 2015, \mn@doi [\apj] {10.1088/0004-637X/799/2/165},
  \href {https://ui.adsabs.harvard.edu/abs/2015ApJ...799..165B} {799, 165}

\bibitem[\protect\citeauthoryear{{Buchner} et~al.,}{{Buchner}
  et~al.}{2014}]{Buchner:2014A&A...564A.125B}
{Buchner} J.,  et~al., 2014, \mn@doi [\aap] {10.1051/0004-6361/201322971},
  \href {https://ui.adsabs.harvard.edu/abs/2014A&A...564A.125B} {564, A125}

\bibitem[\protect\citeauthoryear{{Carretta}, {Bragaglia}, {Gratton}, {D'Orazi}
  \& {Lucatello}}{{Carretta} et~al.}{2009}]{Carretta:2009A&A...508..695C}
{Carretta} E.,  {Bragaglia} A.,  {Gratton} R.,  {D'Orazi} V.,   {Lucatello} S.,
   2009, \mn@doi [\aap] {10.1051/0004-6361/200913003}, \href
  {https://ui.adsabs.harvard.edu/abs/2009A&A...508..695C} {508, 695}

\bibitem[\protect\citeauthoryear{{Clementini} et~al.,}{{Clementini}
  et~al.}{2019}]{gaia_rrl:2019A&A...622A..60C}
{Clementini} G.,  et~al., 2019, \mn@doi [\aap] {10.1051/0004-6361/201833374},
  \href {https://ui.adsabs.harvard.edu/abs/2019A&A...622A..60C} {622, A60}

\bibitem[\protect\citeauthoryear{{Deb}}{{Deb}}{2017}]{debsmc:2017arXiv170703130D}
{Deb} S.,  2017, arXiv e-prints, \href
  {https://ui.adsabs.harvard.edu/abs/2017arXiv170703130D} {p. arXiv:1707.03130}

\bibitem[\protect\citeauthoryear{{Deb}, {Ngeow}, {Kanbur}, {Singh}, {Wysocki}
  \& {Kumar}}{{Deb} et~al.}{2018}]{deblmc:2018MNRAS.478.2526D}
{Deb} S.,  {Ngeow} C.-C.,  {Kanbur} S.~M.,  {Singh} H.~P.,  {Wysocki} D.,
  {Kumar} S.,  2018, \mn@doi [\mnras] {10.1093/mnras/sty1124}, \href
  {https://ui.adsabs.harvard.edu/abs/2018MNRAS.478.2526D} {478, 2526}

\bibitem[\protect\citeauthoryear{{Dierickx} \& {Loeb}}{{Dierickx} \&
  {Loeb}}{2017}]{2017ApJ...836...92D}
{Dierickx} M. I.~P.,  {Loeb} A.,  2017, \mn@doi [\apj]
  {10.3847/1538-4357/836/1/92}, \href
  {https://ui.adsabs.harvard.edu/abs/2017ApJ...836...92D} {836, 92}

\bibitem[\protect\citeauthoryear{Feroz \& Hobson}{Feroz \&
  Hobson}{2008}]{Feroz:2008}
Feroz F.,  Hobson M.~P.,  2008, \mn@doi [Monthly Notices of the Royal
  Astronomical Society] {10.1111/j.1365-2966.2007.12353.x}, 384, 449

\bibitem[\protect\citeauthoryear{Feroz, Hobson  \& Bridges}{Feroz
  et~al.}{2009}]{Feroz:2009}
Feroz F.,  Hobson M.~P.,   Bridges M.,  2009, \mn@doi [Monthly Notices of the
  Royal Astronomical Society] {10.1111/j.1365-2966.2009.14548.x}, 398, 1601

\bibitem[\protect\citeauthoryear{{Frinchaboy}, {Majewski}, {Mu{\~n}oz}, {Law},
  {{\L}okas}, {Kunkel}, {Patterson}  \& {Johnston}}{{Frinchaboy}
  et~al.}{2012}]{Frinchaboy2012ApJ...756...74F}
{Frinchaboy} P.~M.,  {Majewski} S.~R.,  {Mu{\~n}oz} R.~R.,  {Law} D.~R.,
  {{\L}okas} E.~L.,  {Kunkel} W.~E.,  {Patterson} R.~J.,   {Johnston} K.~V.,
  2012, \mn@doi [\apj] {10.1088/0004-637X/756/1/74}, \href
  {http://adsabs.harvard.edu/abs/2012ApJ...756...74F} {756, 74}

\bibitem[\protect\citeauthoryear{{Fritz}, {Battaglia}, {Pawlowski},
  {Kallivayalil}, {van der Marel}, {Sohn}, {Brook}  \& {Besla}}{{Fritz}
  et~al.}{2018}]{Fritz:2018A&A...619A.103F}
{Fritz} T.~K.,  {Battaglia} G.,  {Pawlowski} M.~S.,  {Kallivayalil} N.,  {van
  der Marel} R.,  {Sohn} S.~T.,  {Brook} C.,   {Besla} G.,  2018, \mn@doi
  [\aap] {10.1051/0004-6361/201833343}, \href
  {https://ui.adsabs.harvard.edu/abs/2018A&A...619A.103F} {619, A103}

\bibitem[\protect\citeauthoryear{{Gaia Collaboration} et~al.,}{{Gaia
  Collaboration} et~al.}{2016}]{Gaia_mission}
{Gaia Collaboration} et~al., 2016, \mn@doi [A\&A]
  {10.1051/0004-6361/201629272}, 595, A1

\bibitem[\protect\citeauthoryear{{Gaia Collaboration} et~al.,}{{Gaia
  Collaboration} et~al.}{2018a}]{gaiaDR2:2018A&A...616A...1G}
{Gaia Collaboration} et~al., 2018a, \mn@doi [\aap]
  {10.1051/0004-6361/201833051}, \href
  {https://ui.adsabs.harvard.edu/abs/2018A&A...616A...1G} {616, A1}

\bibitem[\protect\citeauthoryear{{Gaia Collaboration} et~al.,}{{Gaia
  Collaboration} et~al.}{2018b}]{Helmi:2018A&A...616A..12G}
{Gaia Collaboration} et~al., 2018b, \mn@doi [\aap]
  {10.1051/0004-6361/201832698}, \href
  {https://ui.adsabs.harvard.edu/abs/2018A&A...616A..12G} {616, A12}

\bibitem[\protect\citeauthoryear{{Gibbons}, {Belokurov}  \& {Evans}}{{Gibbons}
  et~al.}{2017}]{2017MNRAS.464..794G}
{Gibbons} S.~L.~J.,  {Belokurov} V.,   {Evans} N.~W.,  2017, \mn@doi [\mnras]
  {10.1093/mnras/stw2328}, \href
  {https://ui.adsabs.harvard.edu/abs/2017MNRAS.464..794G} {464, 794}

\bibitem[\protect\citeauthoryear{{Gravity Collaboration} et~al.,}{{Gravity
  Collaboration} et~al.}{2019}]{gravity:2019A&A...625L..10G}
{Gravity Collaboration} et~al., 2019, \mn@doi [\aap]
  {10.1051/0004-6361/201935656}, \href
  {https://ui.adsabs.harvard.edu/abs/2019A&A...625L..10G} {625, L10}

\bibitem[\protect\citeauthoryear{{Hamanowicz} et~al.,}{{Hamanowicz}
  et~al.}{2016}]{hamanowicz:2016AcA....66..197H}
{Hamanowicz} A.,  et~al., 2016, \actaa, \href
  {http://adsabs.harvard.edu/abs/2016AcA....66..197H} {66, 197}

\bibitem[\protect\citeauthoryear{{Hayashi} \& {Chiba}}{{Hayashi} \&
  {Chiba}}{2015}]{2015ApJ...810...22H}
{Hayashi} K.,  {Chiba} M.,  2015, \mn@doi [\apj] {10.1088/0004-637X/810/1/22},
  \href {https://ui.adsabs.harvard.edu/abs/2015ApJ...810...22H} {810, 22}

\bibitem[\protect\citeauthoryear{{Hernitschek} et~al.,}{{Hernitschek}
  et~al.}{2017}]{2017ApJ...850...96H}
{Hernitschek} N.,  et~al., 2017, \mn@doi [\apj] {10.3847/1538-4357/aa960c},
  \href {https://ui.adsabs.harvard.edu/abs/2017ApJ...850...96H} {850, 96}

\bibitem[\protect\citeauthoryear{{Ibata} \& {Lewis}}{{Ibata} \&
  {Lewis}}{1998}]{1998ApJ...500..575I}
{Ibata} R.~A.,  {Lewis} G.~F.,  1998, \mn@doi [\apj] {10.1086/305773}, \href
  {https://ui.adsabs.harvard.edu/abs/1998ApJ...500..575I} {500, 575}

\bibitem[\protect\citeauthoryear{{Ibata}, {Wyse}, {Gilmore}, {Irwin}  \&
  {Suntzeff}}{{Ibata} et~al.}{1997}]{1997AJ....113..634I}
{Ibata} R.~A.,  {Wyse} R. F.~G.,  {Gilmore} G.,  {Irwin} M.~J.,   {Suntzeff}
  N.~B.,  1997, \mn@doi [\aj] {10.1086/118283}, \href
  {https://ui.adsabs.harvard.edu/abs/1997AJ....113..634I} {113, 634}

\bibitem[\protect\citeauthoryear{Iorio \& Belokurov}{Iorio \&
  Belokurov}{2018}]{Iorio:2019}
Iorio G.,  Belokurov V.,  2018, \mn@doi [Monthly Notices of the Royal
  Astronomical Society] {10.1093/mnras/sty2806}, 482, 3868

\bibitem[\protect\citeauthoryear{{Jacyszyn-Dobrzeniecka}
  et~al.,}{{Jacyszyn-Dobrzeniecka}
  et~al.}{2017}]{Jacyszyn-Dobrzeniecka:2017AcA....67....1J}
{Jacyszyn-Dobrzeniecka} A.~M.,  et~al., 2017, \mn@doi [\actaa]
  {10.32023/0001-5237/67.1.1}, \href
  {https://ui.adsabs.harvard.edu/abs/2017AcA....67....1J} {67, 1}

\bibitem[\protect\citeauthoryear{{Jeon}, {Ngeow}  \& {Nemec}}{{Jeon}
  et~al.}{2014}]{jeon:2014IAUS..301..427J}
{Jeon} Y.-B.,  {Ngeow} C.-C.,   {Nemec} J.~M.,  2014, in {Guzik} J.~A.,
  {Chaplin} W.~J.,  {Handler} G.,   {Pigulski} A.,  eds,  IAU Symposium Vol.
  301, Precision Asteroseismology. pp 427--428,
  \mn@doi{10.1017/S1743921313014889}

\bibitem[\protect\citeauthoryear{{Johnston}, {Spergel}  \&
  {Hernquist}}{{Johnston} et~al.}{1995}]{1995ApJ...451..598J}
{Johnston} K.~V.,  {Spergel} D.~N.,   {Hernquist} L.,  1995, \mn@doi [\apj]
  {10.1086/176247}, \href
  {https://ui.adsabs.harvard.edu/abs/1995ApJ...451..598J} {451, 598}

\bibitem[\protect\citeauthoryear{{Jurcsik}}{{Jurcsik}}{1995}]{Jursik:1995AcA....45..653J}
{Jurcsik} J.,  1995, \actaa, \href
  {http://adsabs.harvard.edu/abs/1995AcA....45..653J} {45, 653}

\bibitem[\protect\citeauthoryear{{Kapakos}, {Hatzidimitriou}  \&
  {Soszy{\'n}ski}}{{Kapakos} et~al.}{2011}]{Kapakos:2011MNRAS.415.1366K}
{Kapakos} E.,  {Hatzidimitriou} D.,   {Soszy{\'n}ski} I.,  2011, \mn@doi
  [\mnras] {10.1111/j.1365-2966.2011.18784.x}, \href
  {https://ui.adsabs.harvard.edu/abs/2011MNRAS.415.1366K} {415, 1366}

\bibitem[\protect\citeauthoryear{{Kesden} \& {Kamionkowski}}{{Kesden} \&
  {Kamionkowski}}{2006}]{2006PhRvD..74h3007K}
{Kesden} M.,  {Kamionkowski} M.,  2006, \mn@doi [\prd]
  {10.1103/PhysRevD.74.083007}, \href
  {https://ui.adsabs.harvard.edu/abs/2006PhRvD..74h3007K} {74, 083007}

\bibitem[\protect\citeauthoryear{{Koposov} et~al.,}{{Koposov}
  et~al.}{2012}]{2012ApJ...750...80K}
{Koposov} S.~E.,  et~al., 2012, \mn@doi [\apj] {10.1088/0004-637X/750/1/80},
  \href {https://ui.adsabs.harvard.edu/abs/2012ApJ...750...80K} {750, 80}

\bibitem[\protect\citeauthoryear{{Kuhlen}, {Diemand}  \& {Madau}}{{Kuhlen}
  et~al.}{2007}]{2007ApJ...671.1135K}
{Kuhlen} M.,  {Diemand} J.,   {Madau} P.,  2007, \mn@doi [\apj]
  {10.1086/522878}, \href
  {https://ui.adsabs.harvard.edu/abs/2007ApJ...671.1135K} {671, 1135}

\bibitem[\protect\citeauthoryear{{Law} \& {Majewski}}{{Law} \&
  {Majewski}}{2010}]{2010ApJ...714..229L}
{Law} D.~R.,  {Majewski} S.~R.,  2010, \mn@doi [\apj]
  {10.1088/0004-637X/714/1/229}, \href
  {https://ui.adsabs.harvard.edu/abs/2010ApJ...714..229L} {714, 229}

\bibitem[\protect\citeauthoryear{{{\L}okas}, {Kazantzidis}, {Majewski}, {Law},
  {Mayer}  \& {Frinchaboy}}{{{\L}okas} et~al.}{2010}]{2010ApJ...725.1516L}
{{\L}okas} E.~L.,  {Kazantzidis} S.,  {Majewski} S.~R.,  {Law} D.~R.,  {Mayer}
  L.,   {Frinchaboy} P.~M.,  2010, \mn@doi [\apj]
  {10.1088/0004-637X/725/2/1516}, \href
  {https://ui.adsabs.harvard.edu/abs/2010ApJ...725.1516L} {725, 1516}

\bibitem[\protect\citeauthoryear{{Madore}}{{Madore}}{1976}]{madore:1976RGOB..182..153M}
{Madore} B.~F.,  1976, in The Galaxy and the Local Group. p.~153

\bibitem[\protect\citeauthoryear{{Majewski}, {Skrutskie}, {Weinberg}  \&
  {Ostheimer}}{{Majewski} et~al.}{2003a}]{2003ApJ...599.1082M}
{Majewski} S.~R.,  {Skrutskie} M.~F.,  {Weinberg} M.~D.,   {Ostheimer} J.~C.,
  2003a, \mn@doi [\apj] {10.1086/379504}, \href
  {https://ui.adsabs.harvard.edu/abs/2003ApJ...599.1082M} {599, 1082}

\bibitem[\protect\citeauthoryear{{Majewski}, {Skrutskie}, {Weinberg}  \&
  {Ostheimer}}{{Majewski} et~al.}{2003b}]{Majewski:2003ApJ...599.1082M}
{Majewski} S.~R.,  {Skrutskie} M.~F.,  {Weinberg} M.~D.,   {Ostheimer} J.~C.,
  2003b, \mn@doi [\apj] {10.1086/379504}, \href
  {https://ui.adsabs.harvard.edu/abs/2003ApJ...599.1082M} {599, 1082}

\bibitem[\protect\citeauthoryear{{Majewski} et~al.,}{{Majewski}
  et~al.}{2013}]{Majewski2013ApJ...777L..13M}
{Majewski} S.~R.,  et~al., 2013, \mn@doi [\apjl] {10.1088/2041-8205/777/1/L13},
  \href {http://adsabs.harvard.edu/abs/2013ApJ...777L..13M} {777, L13}

\bibitem[\protect\citeauthoryear{{McConnachie}}{{McConnachie}}{2012}]{mcconnachie:2012AJ....144....4M}
{McConnachie} A.~W.,  2012, \mn@doi [\aj] {10.1088/0004-6256/144/1/4}, \href
  {https://ui.adsabs.harvard.edu/abs/2012AJ....144....4M} {144, 4}

\bibitem[\protect\citeauthoryear{{Millman} \& {Aivazis}}{{Millman} \&
  {Aivazis}}{2011}]{numpy}
{Millman} K.~J.,  {Aivazis} M.,  2011, Computing in Science Engineering, 13, 9

\bibitem[\protect\citeauthoryear{{Nemec}, {Cohen}, {Ripepi}, {Derekas},
  {Moskalik}, {Sesar}, {Chadid}  \& {Bruntt}}{{Nemec}
  et~al.}{2013}]{Nemec:2013ApJ...773..181N}
{Nemec} J.~M.,  {Cohen} J.~G.,  {Ripepi} V.,  {Derekas} A.,  {Moskalik} P.,
  {Sesar} B.,  {Chadid} M.,   {Bruntt} H.,  2013, \mn@doi [\apj]
  {10.1088/0004-637X/773/2/181}, \href
  {https://ui.adsabs.harvard.edu/abs/2013ApJ...773..181N} {773, 181}

\bibitem[\protect\citeauthoryear{{Niederste-Ostholt}, {Belokurov}, {Evans}  \&
  {Pe{\~n}arrubia}}{{Niederste-Ostholt} et~al.}{2010}]{2010ApJ...712..516N}
{Niederste-Ostholt} M.,  {Belokurov} V.,  {Evans} N.~W.,   {Pe{\~n}arrubia} J.,
   2010, \mn@doi [\apj] {10.1088/0004-637X/712/1/516}, \href
  {https://ui.adsabs.harvard.edu/abs/2010ApJ...712..516N} {712, 516}

\bibitem[\protect\citeauthoryear{{Pe{\~n}arrubia} et~al.,}{{Pe{\~n}arrubia}
  et~al.}{2011}]{2011ApJ...727L...2P}
{Pe{\~n}arrubia} J.,  et~al., 2011, \mn@doi [\apjl]
  {10.1088/2041-8205/727/1/L2}, \href
  {https://ui.adsabs.harvard.edu/abs/2011ApJ...727L...2P} {727, L2}

\bibitem[\protect\citeauthoryear{{Price-Whelan} et~al.,}{{Price-Whelan}
  et~al.}{2018}]{astropy:2018}
{Price-Whelan} A.~M.,  et~al., 2018, \mn@doi [\aj] {10.3847/1538-3881/aabc4f},
  \href {https://ui.adsabs.harvard.edu/#abs/2018AJ....156..123T} {156, 123}

\bibitem[\protect\citeauthoryear{Richardson, Spolyar  \& Lehnert}{Richardson
  et~al.}{2014}]{Richardson:2013hga}
Richardson T.,  Spolyar D.,   Lehnert M.,  2014, \mn@doi [Mon. Not. Roy.
  Astron. Soc.] {10.1093/mnras/stu383}, 440, 1680

\bibitem[\protect\citeauthoryear{{Sanders} \& {Evans}}{{Sanders} \&
  {Evans}}{2017}]{Sanders2017MNRAS.472.2670S}
{Sanders} J.~L.,  {Evans} N.~W.,  2017, \mn@doi [\mnras]
  {10.1093/mnras/stx2116}, \href
  {https://ui.adsabs.harvard.edu/abs/2017MNRAS.472.2670S} {472, 2670}

\bibitem[\protect\citeauthoryear{{Sesar}, {Hernitschek}, {Dierickx}, {Fardal}
  \& {Rix}}{{Sesar} et~al.}{2017}]{2017ApJ...844L...4S}
{Sesar} B.,  {Hernitschek} N.,  {Dierickx} M. I.~P.,  {Fardal} M.~A.,   {Rix}
  H.-W.,  2017, \mn@doi [\apjl] {10.3847/2041-8213/aa7c61}, \href
  {https://ui.adsabs.harvard.edu/abs/2017ApJ...844L...4S} {844, L4}

\bibitem[\protect\citeauthoryear{{Skowron} et~al.,}{{Skowron}
  et~al.}{2016}]{Skowron:2016AcA....66..269S}
{Skowron} D.~M.,  et~al., 2016, \actaa, \href
  {https://ui.adsabs.harvard.edu/abs/2016AcA....66..269S} {66, 269}

\bibitem[\protect\citeauthoryear{{Slater} et~al.,}{{Slater}
  et~al.}{2013}]{2013ApJ...762....6S}
{Slater} C.~T.,  et~al., 2013, \mn@doi [\apj] {10.1088/0004-637X/762/1/6},
  \href {https://ui.adsabs.harvard.edu/abs/2013ApJ...762....6S} {762, 6}

\bibitem[\protect\citeauthoryear{{Soszy{\'n}ski} et~al.,}{{Soszy{\'n}ski}
  et~al.}{2014}]{ogleIVRRAB:2014AcA....64..177S}
{Soszy{\'n}ski} I.,  et~al., 2014, \actaa, \href
  {http://adsabs.harvard.edu/abs/2014AcA....64..177S} {64, 177}

\bibitem[\protect\citeauthoryear{{Strigari}, {Frenk}  \& {White}}{{Strigari}
  et~al.}{2010}]{2010MNRAS.408.2364S}
{Strigari} L.~E.,  {Frenk} C.~S.,   {White} S. D.~M.,  2010, \mn@doi [\mnras]
  {10.1111/j.1365-2966.2010.17287.x}, \href
  {https://ui.adsabs.harvard.edu/abs/2010MNRAS.408.2364S} {408, 2364}

\bibitem[\protect\citeauthoryear{{Udalski}, {Szyma{\'n}ski}  \&
  {Szyma{\'n}ski}}{{Udalski} et~al.}{2015}]{ogleIV:2015AcA....65....1U}
{Udalski} A.,  {Szyma{\'n}ski} M.~K.,   {Szyma{\'n}ski} G.,  2015, \actaa,
  \href {https://ui.adsabs.harvard.edu/abs/2015AcA....65....1U} {65, 1}

\bibitem[\protect\citeauthoryear{{Virtanen} et~al.,}{{Virtanen}
  et~al.}{2020}]{2020SciPy-NMeth}
{Virtanen} P.,  et~al., 2020, \mn@doi [Nature Methods]
  {https://doi.org/10.1038/s41592-019-0686-2}, \href {https://rdcu.be/b08Wh}
  {17, 261}

\bibitem[\protect\citeauthoryear{Wickham}{Wickham}{2016}]{ggplot}
Wickham H.,  2016, ggplot2: Elegant Graphics for Data Analysis.
Springer-Verlag New York, \url {https://ggplot2.tidyverse.org}

\bibitem[\protect\citeauthoryear{{Xu} \& {Randall}}{{Xu} \&
  {Randall}}{2019}]{Xu:2019arXiv190408949X}
{Xu} W.~L.,  {Randall} L.,  2019, arXiv e-prints, \href
  {https://ui.adsabs.harvard.edu/abs/2019arXiv190408949X} {p. arXiv:1904.08949}

\bibitem[\protect\citeauthoryear{{van der Marel} \& {Cioni}}{{van der Marel} \&
  {Cioni}}{2001}]{vandermarel:2001AJ....122.1807V}
{van der Marel} R.~P.,  {Cioni} M.-R.~L.,  2001, \mn@doi [\aj]
  {10.1086/323099}, \href
  {https://ui.adsabs.harvard.edu/abs/2001AJ....122.1807V} {122, 1807}

\makeatother
\end{thebibliography}

\appendix 

\section{\emph{Gaia} Queries}\label{appendix:queries}
Selection of all stars in region of the core of Sagittarius:
\begin{verbatim}
SELECT * FROM gaiaDR2.gaia_source
WHERE parallax < 1
	AND ra > 278 AND ra < 290
	AND dec < -28 AND dec > -33
\end{verbatim}
Selection of RRab stars in the region of the core of Sagittarius:
\begin{verbatim}
SELECT * 
FROM gaiaDR2.vari_rrlyrae AS rr 
INNER JOIN gaiaDR2.gaia_source AS gaia 
ON rr.source_id=gaia.source_id
    WHERE parallax < 1
    AND rr.best_classification='RRab'
    AND gaia.ra < 300 AND gaia.ra > 275
    AND gaia.dec < -18 AND gaia.dec > -40
\end{verbatim}{}

\section{Distance derivations for OGLE RR Lyrae}
\label{appendix:distance}
The distances to our RR Lyrae are derived using the same methodology as \citet{Jacyszyn-Dobrzeniecka:2017AcA....67....1J}. The process is briefly summarized and a few important equations are included in this section.
\par Starting from the OGLE-IV bulge catalog all objects with no measurements of the $V$-band magnitude or the Fourier coefficient combination $\phi_{31}^I$ are removed. Next, stars with  atypically small peak-to peak $I$-band amplitudes are excluded (where $A_I < 5\times Log(P)-1$). 
\par For the remaining stars we estimate the metallicity ($\textrm{[Fe/H]}$) of each star photometrically using the period ($P$) and ($\phi_{31}^I$). Following \citet{Skowron:2016AcA....66..269S} the $\phi_{31}^I$ catalog values plus the appropriate $\pi$ offset are converted to $\phi_{31}^V$ (Equation \ref{eq1}), which are then converted to $\phi_{31}^{Kep}$ (\citet{jeon:2014IAUS..301..427J}, Equation \ref{eq2}). Then, the empirical relation from \citet{Nemec:2013ApJ...773..181N} (Equation \ref{eq3}) is used to get ($\textrm{[Fe/H]}$) on the \cite{Jursik:1995AcA....45..653J} scale. 
This metallicity is then converted to the \citet{Carretta:2009A&A...508..695C} scale (\citet{Kapakos:2011MNRAS.415.1366K} Equation \ref{eq4}). Subsequently, the \citet{Braga:2015ApJ...799..165B} Period-Luminosity-Metallicity (PLZ) relation for $W_{I,V-I,abs}$ is applied to derive the absolute Wesenheit magnitude (Equation \ref{eq5}). The observed Wesenheit magnitude \citep{madore:1976RGOB..182..153M} is given by Equation \ref{eq6}. Finally, we use Equation \ref{eq7} to estimate a distance in pc to each RR Lyrae.  

\begin{align}
\phi_{31}^V&=0.122(\phi_{31}^I)^2-0.75(\phi_{31}^I)+5.331 \label{eq1}\\
\phi_{31}^{Kep}&=\phi_{31}^V+0.174 \label{eq2}\\
{\rm [Fe/H]_{J}}&=-8.65-40.12\,P+5.96\,\phi_{31}^{Kep} \label{eq3}\\
                &\quad+6.27\,\phi_{31}^{Kep}\,P-0.72(\phi_{31}^{Kep})^2 \nonumber\\
{\rm [Fe/H]_{C}}&=1.001\,{\rm [Fe/H]_{J}}-0.112 \label{eq4}\\
W_{I,V-I,abs}&=-1.039+-2.524\,Log(P)\label{eq5}\\
             &\quad+0.147 ({\rm [Fe/H]_{C}}+0.04)\nonumber\\
W_{I,V-I}&=I-1.55\,(V-I) \label{eq6} \\
D_{\odot}&=10^{(W_{I,V-I}-W_{I,V-I,abs})/5+1} \label{eq7}
\end{align}

\section{2D projection of a 3d ellipsoid}\label{appendix:projprop}
Initially our ellipsoid can be described in its inertial frame by the following equation $\frac{x^2}{a^2}+\frac{y^2}{b^2}+\frac{z^2}{c^2}=1$.
In matrix form:
\begin{equation*}
E
=
\begin{bmatrix}
a^{-2} & 0 & 0\\
0 & b^{-2} & 0\\
0 & 0 & c^{-2}
\end{bmatrix}
\end{equation*}
and
\begin{equation*}
\begin{bmatrix}
x & y & z
\end{bmatrix}{}
E
\begin{bmatrix}
x\\
y\\
z\\
\end{bmatrix}
=1.
\end{equation*}
we then rotate this ellipse into the observed frame see Equation \ref{eqn:rotation}. 
\begin{align}
T(\gamma,\pi/2-\phi,\theta)^{\rm T}
&\begin{bmatrix}
a^{-2} & 0 & 0\\
0 & b^{-2} & 0\\
0 & 0 & c^{-2}
\end{bmatrix}
T(\gamma,\pi/2-\phi,\theta)
=\nonumber \\
&\begin{bmatrix}
A_{11} & A_{12} & A_{13}\\
A_{21} & A_{22} & A_{23}\\
A_{31} & A_{32} & A_{33}
\end{bmatrix}
=1
\end{align}
The conic section equation is given by:
\begin{align}\label{eqn:conicsection1}
f(x',y',z')=&A_{11} x'^{2}+ A_{22}y'^{2} +A_{33}z'^{2}\nonumber\\
&+2 A_{12}x'y' + 2 A_{13}x'z'+ 2 A_{23}y'z'\\=&1\nonumber
\end{align}
The ellipse formed by the \emph{shadow} of this ellipsoid on the observed $x'y'$ plane is defined as the set of points where the $z'$ component of $\nabla f(x',y',z')=0$. 
\begin{align*}
    \frac{\dd f}{\dd z}&=2 A_{33}z'+2 A_{13}x'+2 A_{23}y'=0\\
    z'&=\frac{-A_{13}x'-A_{23}y'}{A_{33}}
\end{align*}
plugging this into equation \ref{eqn:conicsection1} gives the conic section equation
\begin{align*}
    1&=A_{11} x'^{2}+ A_{22}y'^{2} +A_{33}\bigg(\frac{-A_{13}x'- A_{23}y'}{A_{33}}\bigg)^{2}+2 A_{12}x'y'\\
    &+ 2 A_{13}\frac{- A_{13}x'- A_{23}y'}{ A_{33}}x'+ 2 A_{23}\frac{- A_{13}x'- A_{23}y'}{ A_{33}}y'\\
\end{align*}
Grouping like terms gives

\begin{align*}
    0=-1+\bigg(A_{11}-\frac{A_{13}^2}{A_{33}}\bigg) &x'^2\\
     +\bigg(2A_{12}-2\frac{A_{13}A_{23}}{A_{33}}\bigg) &x'y'\\
     +\bigg(A_{22}-\frac{A_{23}^2}{A_{33}}\bigg) & y'^2
\end{align*}
This is the canonical conic section equation with D \& E equal to 0.
\begin{equation*}
    Ax'^2+Bx'y'+Cy'^2+Dx+Ey+F=0
\end{equation*}
We then define a normalization factor (K) and the linear eccentricity or the distance of the focus from the center of the ellipse (s).
\begin{align}
    K&=64\,F\,(4\,A\,C-B^2)/(4\,A\,C-B^2)^2\\
    s&=\frac{1}{4}\sqrt{\given K\given\,\sqrt{B^2+(A-C)^2}}
\end{align}{}
From here we can write relations for the semi-major ($a_{\textrm{proj}}$) and semi-minor ($b_{\textrm{proj}}$) axes.
\begin{align}
        a_{\textrm{proj}}&=\frac{1}{8}\sqrt{2|K|\sqrt{B^2+(A-C)^2-2q(A+C)}}\\
        b_{\textrm{proj}}&=\sqrt{a_{\textrm{proj}}^2-s^2}
\end{align}{}
and the projected position angle ($P.A.$) is given by 
\begin{equation}
    P.A.=\frac{3 \pi}{2}-\frac{1}{2} \textrm{atan2}\,\bigg(\frac{b}{a-c}\bigg) 
\end{equation}

\section{Joint Posterior corner plot}\label{appendix:corner} 
\par
\begin{figure*}
    \centering
    \includegraphics[width=0.98\textwidth]{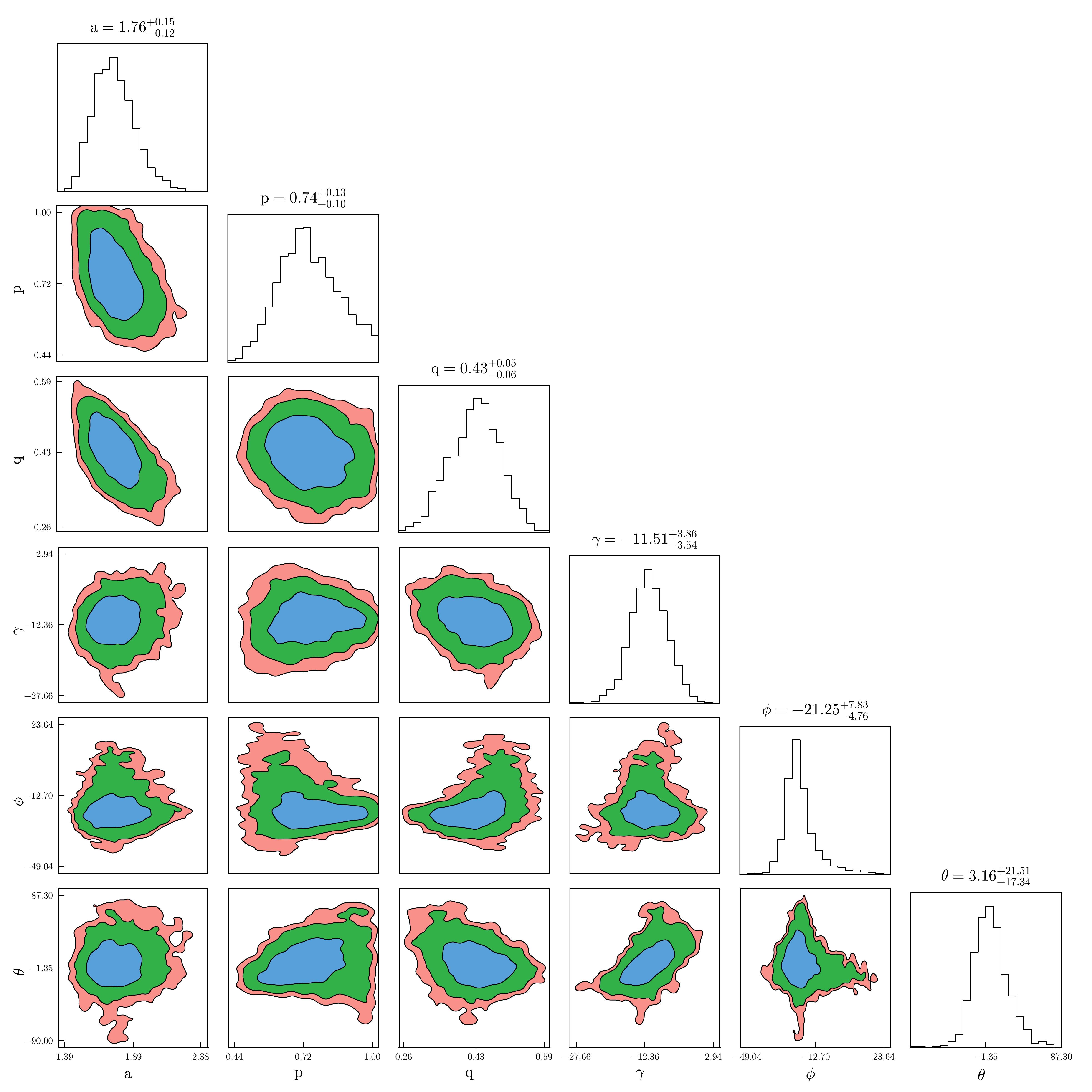}
    \caption{Posterior probability densities from a joint fit to the \emph{Gaia} and OGLE data for the six baseline parameters of our model; $a$ is the major axis scale length in $kpc$, $p$ is the ratio of the intermediate to the major axis, and $q$ is the ratio of the minor to the major axis. The angles are the Euler rotation angles as defined in Section~\ref{sec:Methods} in units of degrees.
    \label{fig:jointcorner}}
\end{figure*}

 Figure~\ref{fig:jointcorner} shows the posterior probability densities for our baseline set of six model parameters from a joint fit to the \emph{Gaia} and OGLE  data. 
{}
\bsp	
\label{lastpage}
\end{document}